\documentstyle[11pt,aaspp4]{article}

\lefthead{Calzetti \& Heckman}
\righthead{Opacity of Galaxies}

\begin{document}

\title{The Evolution of Dust Opacity in Galaxies.}

\author{Daniela Calzetti}
\affil{Space Telescope Science Institute, 3700 San Martin Dr., 
   Baltimore, MD 21218, USA; e-mail: calzetti@stsci.edu}
\and
\author{Timothy M. Heckman\altaffilmark{1}}
\affil{Dept. of Physics and Astronomy, The Johns Hopkins University, 
Baltimore, MD 21218, USA; e-mail: heckman@pha.jhu.edu}

\altaffiltext{1}{Adjunt Astronomer at the Space Telescope Science Institute.} 

\begin{abstract}
We investigate the evolution of the opacity of galaxies as a function
of redshift, using simple assumptions about the metal and dust
enrichment of the gas and the distribution of dust in galaxies. The
method presented in this paper uses an iterative procedure to
reconstruct the intrinsic Star Formation Rate (SFR) density of
galaxies as a function of redshift, by correcting the observed UV
emission for the obscuring effects of the dust produced by the star
formation itself. Where necessary, gas inflows/outflows are included
to dilute/remove the excess of metals produced by the stars for fixed
yield. The iterative procedure converges to multiple solutions for the
intrinsic SFR density, divided into two basic classes.  The first
class of solutions predicts a relatively large amount of UV
attenuation at high redshift, with A$_{1500\AA}$=1.9~mag at z$\sim$3,
and smaller attenuations at low redshift (z$<$1), with
A$_{2800\AA}$=1.25~mag. The SFR density of this set of solutions is
constant for z$\gtrsim$1.2 and declines for z$<$1.2; it resembles in
shape the ``monolithic collapse'' scenario for star formation.  The
second class of solutions predicts relatively low UV attenuations at
high redshift, with A$_{1500\AA}$=0.75~mag at z$\sim$3, and larger
attenuations at low redshift, with A$_{2800\AA}$=1.50~mag at z$<$1.
The SFR density in this case has a peak at z$\simeq$1.2. The
advantages and shortcomings of both classes are analyzed in the light
of available observational constraints, including the opacity of
galaxies at 0$\le$z$\le$1 and the intensity and spectral energy
distribution of the cosmic infrared background from the COBE DIRBE and
FIRAS data. We conclude that both classes of models are acceptable
within the current uncertainties, but the ``monolithic collapse''
class matches the available observations better than the other one. We
also investigate the dependence of our solutions on the different
model assumptions.
\end{abstract}

\keywords{Galaxies: Evolution -- Galaxies: ISM -- ISM: Dust, Extinction }

\section{Introduction}

One of the fundamental elements of the study of the evolution of the
Universe is the investigation of how the light evolves. The bulk of
the observed light is given by the stellar component of galaxies, and
measuring the light and its spectral energy distribution measures the
total stellar content and the star formation history of
galaxies. Using ultraviolet (UV) data from the Canada-France Redshift
Survey (Lilly et al. 1996) and from the HDF (Williams et al. 1996),
Madau et al. (1996) estimated the cosmic star formation rate density of
galaxies. The UV emission is more directly related to the star
formation rate (SFR) than emission at longer wavelengths, because most
of the UV light below 3,000~\AA~ is emitted by hot, massive,
relatively short-lived ($<$1~Gyr) stars. In addition, the rest frame
UV is the most readily accessible wavelength range in high redshift
galaxies with standard (optical) observing techniques.

One of the problems which haunts the UV determinations of the cosmic
SFR density is the potentially large uncertainty introduced by dust
obscuration.  Star formation comes together with metal production and,
therefore, dust. The timescales for the formation of dust are short,
$\sim$100-200~Myr (Tielens 1990, Jones et al. 1994, Dwek 1998), and
dust opacity may affect the emerging light fairly early in the history
of a galaxy.  Dust extinction is more effective at shorter
wavelengths, so the rest frame UV light is more affected than the
optical or IR emission. The degree to which the light is affected is,
however, difficult to quantify as the metal (and dust) content per
unit mass is only one of the parameters to take into account; for
instance, the opacity of a galaxy is also determined by its gas
fraction. Although present-day galaxies are more metal rich than high
redshift galaxies, they could be less opaque; they have locked a large
fraction of their gas into stars, thus removing dust from the
interstellar medium. In the past, more gas was available per unit
mass; thus even small metallicities had potentially large effects in
terms of the obscuring power of the associated dust. Last, but not
least, is the uncertainty introduced by the unknown distribution of
the dust in the galaxy. The effects on the emerging UV-optical light
of various geometries of emitters and absorbers have been discussed by
a number of authors (to name a few, Rowan-Robinson \& Crawford 1989,
Witt, Thronson \& Capuano 1992, Calzetti, Kinney \& Storchi-Bergmann
1994, Witt \& Gordon 1996), and this is the largest uncertainty in any
model of galaxy opacity.

The problem of dust obscuration in the derivation of the cosmic SFR density is 
widely recognized and different authors have taken different
approaches to it. Pei \& Fall (1995) introduce dust in their treatment
of the evolution of Damped Lyman-$\alpha$ systems (DLAs) to account
for missing high metallicity systems in high redshift surveys.  Meurer
et al. (1997) have been among the first to raise the issue for
Lyman-break galaxies, by noticing that the UV spectral energy
distributions of these objects (Steidel et al. 1996) are too red to be
dust-free star-forming galaxies. They attempt a correction for dust
reddening of the high redshift galaxies by extrapolating a method used
for local starburst galaxies (Calzetti et al.  1994). Sawicki \& Yee
(1998) reach similar conclusions using a wider photometric baseline.
Rowan-Robinson et al. (1997) use ISO observations to derive the star
formation rate at high redshift from the far-infrared emission of
galaxies.  Renzini (1997) uses evidence on the metal enrichment of the
intracluster medium (see, e.g., Yamashita 1992) to infer the amount of
high redshift star formation required to produce the observed mass in
metals. Mushotzky \& Loewenstein (1997) use the lack of evolution in
the metal abundance of galaxy clusters to conclude that most of the
metal enrichment occurred at z$>1$ and, thus, star formation had to be
higher than measured from UV-optical surveys. Madau, Pozzetti, \&
Dickinson (1998) discuss the case of a simple model of foreground dust
with increasing opacity as a function of redshift, imposing negligible
dust correction at low redshift and a factor $\sim$5 in the UV flux at
z$\sim$3. In all cases, there is no attempt to obtain a
self-consistent solution between the correction factor imposed to the
global SFR density (derived from the UV light of galaxies) and the UV opacity
such SFR density implies.

There is mounting evidence that galaxies in the local Universe are
generally not very opaque, although the quantification of this
statement is still under debate. With techniques which use
illuminating background sources (e.g.: White, Keel \& Conselice 1996,
Berlind et al. 1997, Gonzalez et al. 1998), spiral galaxies are shown
to have transparent interarm regions at optical wavelengths, with
A$_B\sim$A$_I\sim$0.1-0.2, and opaque arms, with
A$_B\sim$A$_I\sim$1. The same level of opacity as the arms is found in
the centers of the spirals (Giovanelli et al. 1995, Moriondo,
Giovanelli \& Haynes 1998). As expected, the opaque regions of spirals
coincide with the regions of most intense star formation (the arms),
as stars form in the dusty molecular clouds. The UV-optical energy
absorbed by dust is re-radiated in the FIR at wavelengths longer than
a few $\mu$m.  The energy balance between the UV-optical and the
Far-Infrared (FIR) emission from disk galaxies reveals that the amount
of stellar light absorbed by dust and re-emitted in the FIR range is
about the same as the emerging stellar light (Soifer \& Neugebauer
1991, Xu \& Buat 1995, Wang \& Heckman 1996, Trewhella 1997).
Preliminary results from ISO observations of the CFRS galaxies
indicate that about 55\%~ of the star formation at z$<$1 emerges in
the FIR (Hammer \& Flores 1998). Local (z$\le$0.1) surveys in
H$\alpha$ show that the SFR derived from the nebular emission line
luminosity density is about a factor $\sim$3 higher than the low
redshift extrapolation of the CFRS UV measurements (Gronwall 1998,
see, however, Gallego et al. 1995), suggesting that the lower value of
SFR from the UV may be due to dust obscuration.  The low-redshift
issue, however, is not settled: Treyer et al. (1998) find that their
UV data from FOCA also produce a local (z$\sim$0.15) value of the SFR
which is a factor $\sim$2--3 larger than the extrapolation from the
CFRS UV data. Although none of the above results can be considered
final, the available data suggest that dust absorbs a factor
$\sim$1/2, and probably no more than 2/3, of the total stellar light
in local galaxies.

The situation is even less clear at high redshift, as long wavelength
baselines are available only for a handful of high redshift
galaxies. The optical and near-IR rest frame of distant galaxies is in
the near-mid IR range, and ground-based measurements are less
sensitive at these wavelengths. On the basis of the UV slope alone,
Meurer et al. (1997) suggest a factor 10 attenuation of the 1500~\AA~
flux of the Lyman-break galaxies at z$\sim$3, although recent
revisions bring the attenuation correction down to 5.5 (Meurer,
Heckman \& Calzetti 1998). The addition of near-IR data points brings
the 1500~\AA~ attenuation at z$\sim$3 into the factor range 2.5--7
(Dickinson 1998). The ISO FIR emission of high redshift galaxies in
the Hubble Deep Field indicates a correction factor closer to $\sim$5
at z$>2$ (Rowan-Robinson et al. 1997). The nebular line emission of
six of the Lyman-break galaxies from near-IR spectroscopy also suggest
corrections between 3 and 6 for the flux at 1500~\AA~ (Pettini et
al. 1998a). Optical/near-IR nebular line
ratios and CO detections have provided additional observational
evidence for dust at redshifts as high as z=4.70 in several individual
objects, usually radio galaxies (e.g., Dey, Spinrad \& Dickinson 1995,
Armus et al. 1998a) and AGNs (e.g., Omont et al. 1996, Ivison et
al. 1998). Also, indirect evidence for presence of dust in a z=5.34
galaxy has been presented by Armus et al. (1998b).

Current measurements of the cosmic infrared background (CIB) from COBE
DIRBE (Hauser et al. 1998) and FIRAS (Fixsen et al. 1998) imply a
non-negligible FIR emission from dust in galaxies. For instance, the
DIRBE data give $\nu$I($\nu$)=25$\pm$7~nW~m$^{-2}$~sr$^{-1}$ at
140~$\mu$m and 14$\pm$3~nW~m$^{-2}$~sr$^{-1}$ at 240~$\mu$m, which
imply a CIB emission between 2 and 3 times higher that the combined
emission of the UV$+$optical$+$near-IR backgrounds (Madau et
al. 1998). Still unknown is the mean redshift of the galaxies from
which the bulk of the CIB arises. Recent observations obtained at
450~$\mu$m and 850~$\mu$m with SCUBA at the JCMT (Smail, Ivison \&
Blain 1997, Hughes et al. 1998, Barger et al. 1998, Eales et al. 1998)
have attempted to address this issue by probing the FIR emission of
individual high redshift galaxies. The SCUBA data should provide a
measure of the intrinsic, reddening-free SFR in high redshift galaxies
as much as the IRAS data provide a measure of the intrinsic SFR in
local starburst galaxies (e.g., Young et al. 1989). The SCUBA
detections are consistent with many of the objects being at z$\ge$1;
the inferred SFRs have values $\sim$2--5 times higher than those
derived from the restframe UV detections in the same redshift
range. However, in the local Universe the brightest FIR emitters are
generally AGNs or powerful starbursts, like Arp220, and the nature of
the SCUBA galaxies (AGN/starburst) has not been settled yet. In
addition difficulties in the redshift assignment leave some of the
interpretation still open for debate. The flux density of the current
SCUBA sources amounts to about 20--30\% of the COBE background at
850~$\mu$m, and the number density is only 1/10 of the optical high
redshift galaxies (Lilly et al. 1998).

The aim of this paper is to model the average opacity of galaxies as a
function of both redshift and wavelength. We mainly concentrate on the
effects of dust on the ensemble of restframe UV photons produced by
galaxies throughout cosmic times. The basic philosophy is the
following. Star formation produces metals and, thus, dust. The global
SFR density of galaxies is derived from the observed UV luminosities,
which are sensitive to dust opacity. We solve self-consistently for
the intrinsic global SFR density and for the effects of the dust
created by the star formation itself, till we reproduce the observed
UV flux density. Simple assumptions about the cosmology, and the
galaxy and dust models allow us to budget the fraction of light lost
to dust obscuration at each redshift. From the computed opacities, we
then derive the intensity and the spectral energy distribution of the
CIB, which we compare with the observational data from COBE DIRBE and
FIRAS. Additional observational constraints, such as DLAs and SCUBA
measurements at high redshift, and the energy density in the local
Universe at FIR (both IRAS and ISO), H$\alpha$, and radio wavelengths
are used to further validate the solutions. Finally, we discuss the impact
of our input assumptions on the results.

In the following, we adopt a cosmology with H$_o$=50~km/s/Mpc and
q$_o$=0.5.

\section{The Model}

For standard assumptions as to the cosmology, the metal and dust
enrichment, and the distribution of dust in galaxies, we follow in
time the evolution and appearance of the average galaxy from the
beginning of the star formation to the present. Here galaxies are
assumed to be already gravitationally bound, although not closed,
systems. Only the evolution of the stellar, metal, and dust contents
is implemented. Edmunds \& Phillips (1997) present an extensive
treatment of the chemical evolution of galaxies for different
assumptions about the star formation history and galaxy model
(closed-box/inflow/outflow). Their work is in the same vein as Pei \&
Fall (1995), who discuss the evolution of DLAs. The main difference
between our and those authors' approach is that we do not adopt an
{\it a priori} global SFR density; rather, we start from the observed one, 
derived from the UV flux density of galaxies, and
correct it iteratively for the associated UV dust opacities, until the
corrections match the predicted obscuration. The observed global SFR density 
predicts non-zero UV dust opacity from the simple act of forming
stars. These opacities are fed back into the SFR density in the form of
correction factors; the evolution of the star, metal and dust content
of galaxies is followed again and new opacities derived. The iterative
procedure is stopped when the corrections to the global SFR density differ
1~$\sigma$ or less from the UV opacities, where the 1~$\sigma$ is from
the observational uncertainties. Finally, we use the derived dust
opacities to infer the intensity and spectral energy distribution of
the CIB.

A general characteristic of the model is that galaxies are not a
priori closed boxes, but can be experiencing outflows of
metal-enriched gas or inflows of metal-free gas. The details of the
evolution of each galaxy (e.g., merging, see White \& Frenk 1991,
versus monolithic collapse, see Ortolani et al. 1995) has little
influence on the model, since the computation of the metal and dust
fractions does not require assumptions as to the specific
characteristics of each galaxy and its interaction with the
environment (e.g. presence/absence of merging). To some extent, the
infall of small substructures into more massive systems, as predicted in
scenarios of hierarchical assembly of galaxies (e.g.,
Aragon-Salamanca, Baugh \& Kauffmann 1998), can be considered similar
to gas inflows, although differences may exist as the infalling
fragments do not necessarily contain pristine gas. The determination
of the dust opacity requires, however, assumptions on the galaxy's
mass distribution and on the dust distribution relative to the stars,
which we will derive entirely from observational constraints.

\subsection{Inputs to the Model}

\subsubsection{The Baryonic Mass in Galaxies}

The total mass locked today in galaxies as stars, planets, dust and
gas is calculated to be $\Omega_{star+gas}\simeq 0.0059$~h$_{50}^{-1}$
(Gnedin \& Ostriker 1992, Fukugita, Hogan \& Peebles 1998), or about
10\%~ of the total mass in baryons (Schramm \& Turner 1998). Similar
numbers can be directly obtained from the local galaxy luminosity
function (Marzke, Huchra \& Geller 1994). For
$\phi_*$=0.005~h$_{50}^{3}$ and
L$^*_B$=2.4$\times$10$^{10}$~h$_{50}^{-2}$~L$_{\odot}$, the luminosity
density at B in the local Universe is 
L$_B\simeq$1.2$\times$10$^8$~h$_{50}$~L$_{\odot}$~Mpc$^{-3}$; with an
average mass-to-light ratio for the galaxies M/L$_B$=3.4, the average
mass density in co-moving space is
M$_{tot}$=4.08$\times$10$^8$~h$_{50}$~M$_{\odot}$~Mpc$^{-3}$, which is
equivalent to the $\Omega_{star+gas}$ value given above. The average
M/L$_B$ value used in the above calculation is obtained as a weighted
mean, assuming that 20\% of the galaxies are Ellipticals, with
M/L$_B$=10 (Worthey 1994), 15\% are S0 with M/L$_B$=5 and the
remaining 65\% are spirals and irregulars with M/L$_B$=1.

Kennicutt (1998) measures a typical gas surface density
$<\Sigma_{gas}>\simeq 15$~M$_{\odot}$~pc$^{-2}$ in local
spirals. Since the fraction of residual gas in these systems is about
15\% of the total baryonic mass, the average surface mass density
$<\Sigma_{star+gas}>\simeq 100$~M$_{\odot}$~pc$^{-2}$.  For the
adopted luminosity function and mass-to-light ratio, the average
galaxy has total baryonic mass
M$^*$=8.2$\times$10$^{10}$~h$_{50}^{-2}$~M$_{\odot}$, and the average
surface mass density of 100~M$_{\odot}$~pc$^{-2}$ implies a radius of
16~h$_{50}^{-1}$~kpc. The resulting virial velocity from this baryon
component is $\sim$150~km/s. For comparison, in our Milky Way the
rotation velocity due to the baryonic component only (disk plus bulge)
within its optical radius, $\sim$12~kpc, is between 120 and
160~km~s$^{-1}$, versus a total rotation velocity between 200 and
230~km~s$^{-1}$ due to the combined effect of the baryonic matter and the dark
matter halo (Dehnen \& Binney 1998).

The local $\Omega_{star+gas}$ is uncertain by at least 50\%~ (Gnedin
\& Ostriker 1992, Fukugita et al.  1998), and is one of the parameters
in the model with the largest uncertainty. $\Omega_{star+gas}$ is a
lower limit to $\Omega_{gas}(z=\infty)$ if galaxies undergo gas
outflows and, thus, were more massive in the past than today, or it is
an upper limit if galaxies experience gas inflows. We note that the
presence of metals in the intracluster medium (e.g. Yamashita 1992)
strongly favors outflows from galaxies, or a combination of outflows
and inflows.

\subsubsection{The Stellar Initial Mass Function}

The shape, mass limits, and universality of the stellar Initial Mass
Function (IMF) are still matters of heated debate, and choosing an IMF
for a model is far from being free of problems. But this is also one
of the most crucial steps in the process. The SFR history is presently
derived from the measured UV emission of galaxies; the UV emission,
however, probes only the light from the most massive stars, and the
extrapolation to a SFR density requires adopting an IMF. Most of the mass is
locked in stars with mass below $\approx$1--2~M$_{\odot}$; thus the
low-mass end of the IMF is the main determinant of the SFR value 
and, therefore, of the gas consumption rate in
galaxies. 

Derivations of the high-mass end of the IMF in star-forming regions
of nearby galaxies indicate a Salpeter slope above
$\sim$5--10~M$_{\odot}$ and an upper limit of 100~M$_{\odot}$ (Massey
et al. 1995, Hunter et al. 1996, 1997), independent of environmental
factors, such as metallicity, galaxy type or location within the
galaxy. Contrary to the high-mass end, the shape and low mass limit of
the IMF below $\approx$1--2~M$_{\odot}$ is the least known part of the
function (see the reviews of Scalo 1998 and Larson 1998). Since most
of the light is provided by the most massive stars, direct observations of
low mass stars are extremely difficult. Measurements of the IMF in the
solar neighborough and in globular clusters indicate the IMF peaks
between 0.2 and 0.5~M$_{\odot}$ and has a sharp decline in the number
of stars produced below this mass (Kroupa 1995, Scalo 1998).  A recent
measurement in R136, the central young stellar cluster in the 30~Dor region 
of the Large Magellanic Cloud, shows that the IMF is Salpeter-like
($\alpha$=2.35) down to $\sim$2~M$_{\odot}$, and flattens ($\alpha$=1)
below that mass (Sirianni et al. 1998). Current uncertainties in the
low-mass end of the stellar IMF imply uncertainties in the SFR history
up to factors 2--3.

Following the above results, we adopt a Salpeter IMF with a
lower cut-off at 0.35~M$_{\odot}$ and an upper cut-off at
100~M$_{\odot}$.  No stars are produced below 0.35~M$_{\odot}$. The
number of stars generated is a factor 1.64 less than for a
0.1--100~M$_{\odot}$ Salpeter IMF, and a factor 1.55 more than for a
1--100~M$_{\odot}$ Salpeter IMF.  All stars above 8~M$_{\odot}$
explode as supernovae, thus recycling metal-rich gas into the ISM. For
our IMF, the return fraction due to supernovae is R=0.23. For star
formation extended over timescales of 1~Gyr or more, the fraction of
gas returned to the ISM must include the contribution from stars
between 1 and 8~~M$_{\odot}$: these stars provide a delayed return of
50\%~ to 80\%~ of the initial mass in form of stellar winds and
planetary nebula ejecta. Kennicutt, Tamblyn \& Congdon (1994)
calculated the contribution of these low-mass stars; their results,
normalized to our IMF, imply an effective return fraction
R$_{eff}$=0.46. Thus, the amount of gas permanently locked into stars
over Gyr timescales is (1$-$R$_{eff}$)=0.54 of the total formed and
this effective value is what we adopt in our computation.

\subsubsection{The Star Formation History}

The observed UV emission of galaxies provides the starting point for
calculating the history of the SFR density, as the UV light probes the most
massive stars and is thus a measure of recent star formation. The UV
emission of galaxies has been estimated in the redshift range
z$\simeq$0--4 (Lilly et al. 1996, Madau et al. 1996, Connolly et
al. 1997). Once an average star formation history and an IMF are
assumed for the galaxies, the UV luminosity can be converted into a
star formation rate as a function of redshift, SFR(z) (cf. Madau et
al.  1996). The observed UV emission corresponds to different
rest-frame wavelengths at different redshifts: $\sim$2800~\AA~ in the
range 0$\le$z$\le$2 and $\sim$1500~\AA~ for z$>2.5$. For the adopted
IMF, which is truncated at the lower end at 0.35~M$_{\odot}$, a galaxy
which is forming stars at a constant rate of 1~M$_{\odot}$/yr produces
1.475$\times$10$^{28}$~erg~s$^{-1}$~Hz$^{-1}$ in the UV, with little
variation between 1,500~\AA~ and 2,800~\AA. The observed SFR(z),
SFR$_{obs}$(z), is a {\it lower limit} to the intrinsic star
formation, SFR$_{int}$(z), because of the effects of dust obscuration
mentioned earlier.

A source of uncertainty for the conversion from UV emission to SFR(z)
is the dependence of the UV emission on the star formation history of
the galaxies, especially at longer wavelengths. The flux at
$\lambda<$2000~\AA~ is due predominantly to the short-lived
M$>$10~M$_{\odot}$ stars and is, therefore, a reasonable estimator of
the recent SFR (for a fixed IMF). On the contrary, the UV flux above
2000~\AA~ probes stars with lower and lower masses and becomes more
and more sensitive to the star formation history of the galaxy. For
instance, F($\nu_1$)/F($\nu_2$)$\simeq$1 for the 1500~\AA~ to 2800~\AA~
ratio for constant star formation, F($\nu_1$)/F($\nu_2$)$\simeq$1.5
for a 1~Myr old instantaneous burst and F($\nu_1$)/F($\nu_2$)$\simeq$1
for a 15~Myr old instantaneous burst. The difference is exacerbated
if the burst is lying on top of a pre-existing stellar population
(e.g., the residual from previous bursts). Uncertainties in the star
formation history, therefore, account for about 30--50\%~ variation 
(and possibly more) in the conversion UV-to-SFR(z).

Another source of uncertainty is the potential underestimate of the
global UV flux due to incomplete sampling of the lower end of the
galaxy luminosity function and, even worse, to uncertain volume
corrections. This uncertainty will predominantly affect the high
redshift data points, and the point at z=4 in Madau et al. more than
the point at z=2.75. The volume corrections for the UV flux density at
z=2.75 are derived from a sample of $\sim$500 Lyman-break galaxies
with confirmed redshifts (e.g., Steidel et al. 1998, Dickinson 1998;
Giavalisco 1998, private communication); thus, for all purposes, we
can consider this point as only slightly affected by
incompleteness. The UV flux density at z=4, on the contrary, has been
entirely derived from photometric redshifts (B-band dropouts) in two
relatively small fields; the z=4 candidate galaxies have not been
spectroscopically confirmed yet (except for a handful of recent
detections, Dickinson 1998, private communication), so the volume
corrections in this case are highly uncertain. Thus, except in one
case (see below), we will not attempt to use the z=4 data point as a
constraint to the solutions.

\subsubsection{Metals and Dust Production}

Once stars start forming, metal production follows with a negligible
time delay, as the first supernovae explode after only 3--4~Myr from
the onset of star formation (Leitherer \& Heckman 1995). Our emphasis
is on the dust content rather than the chemistry  galaxies;
therefore we adopt here the simplest metal-enrichment scenario, the
instantaneous mixing model, to derive metallicities. This model should
be adequate to follow the global metal enrichment of galaxies,
although it may be insufficient in the case of individual objects
(see, e.g., Kunth, Matteucci \& Marconi 1995; Tosi 1998). For
instantaneous mixing, the relation between the final metallicity of
the gas and the fraction of residual gas in the galaxy is given by 
(Pei \& Fall 1995):
\begin{equation}
Z_{gas}(z)=-{\alpha_Z\over 1+\gamma}  ln[M_{gas}(z)/M_{tot}(\infty)];
\end{equation}
where M$_{gas}$(z) is the galaxy mass in gas at redshift z,
M$_{tot}(\infty)$ is the total mass in galaxies at the beginning of
star formation, $\alpha_Z$ is the true yield of the stars, and
$\gamma$ is the outflow parameter; $\gamma >$0 implies gas outflows,
while $\gamma$=0 implies a closed box. The analogous formula for gas
inflows is given in Pei \& Fall (1995) and is not repeated here.  The
true yield $\alpha_Z$ is strongly dependent on the stellar IMF
(e.g. Pagel 1987, 1998). Edmunds \& Phillips (1997) measure for the
global oxygen abundance of galaxies [O/H]/[O/H]$_{\odot}\sim-$0.1,
which implies a yield $\alpha_Z\sim$0.7--0.8 or higher. We adopt
$\alpha_Z$=0.7 (cf. Table~1); this is an intermediate value between
the extremes observed in our Galaxy (Pagel 1987), but we bear in mind
that it could be a lower limit, and is uncertain by a factor
$\sim$2. This is another difference between our and Edmunds \&
Phillips' (1997) and Pei \& Fall's (1995) approach. We adopt a fixed
value for the yield and change the inflow/outflow parameters; the
other authors change the yield according to a preselected
outflow/inflow parameter. In both cases, however, the target is to
contrain the model galaxies to have a predetermined value of the
metallicity at z=0. Equation~1 follows from the assumption that the
mass of ejected/accreted gas is proportional to the SFR: $\delta
M_{out/in}(z)/\delta t(z) = \pm \gamma$~(1$-$R$_{eff}$)~SFR(z), with
$t(z)$ the time corresponding to redshift $z$ (Pei \& Fall 1995).  The
final metallicity of the gas at zero redshift is here constrained to be
Z$_{gas}$(0)=Z$_{\odot}$, the mean gas metallicity of the average
L$^*$ galaxy, with the fraction of residual gas being the observed
value of $\sim$15\%~ (Fukugita et al. 1998).

The initial mass of gas bound in galaxies at z=$\infty$,
M$_{tot}$($\infty$), is given by the value
M$_{tot}$(0)=4.08$\times$10$^8$~h$_{50}$~M$_{\odot}$~Mpc$^{-3}$
augmented/decreased by the amount of material ejected into/accreted
from the IGM over the history of the galaxy.  Thus, the initial mass
is given by the formula:
\begin{equation}
M_{tot}(\infty) = M_{tot}(0) \pm {\gamma} M_{star}(0),
\end{equation}
where M$_{star}(0)$ is the total mass in stars at z=0. As will be
discussed in the following sections, a typical galaxy can be expected
to eject $\sim$40\%~ or accrete $\sim$50\%~ of its initial mass
into/from the intergalactic medium (IGM).

Once metals are produced, dust follows within $\sim$100-200~Myr (e.g.,
Tielens 1990, Jones et al. 1994, Dwek 1998). Over the same timescale
the dust/metal ratio comes to within a factor $\sim$2 of its
equilibrium value (Dwek 1998). Studies of local galaxies indicate that
the dust/gas mass ratio increases with the metallicity of the gas
(Bouchet et al. 1985). Observations of DLAs at redshift z$\sim$2--3,
in addition, suggest that the relation between dust/gas and
metallicity may not be linear, namely that the conversion from metals
to dust may be itself a function of metallicity. In DLAs,
(dust/metals)$_{DLAs}$=0.5(dust/metals)$_{MW}$ at
Z$_{gas}\sim$1/15~Z$_{\odot}$ (Pettini et al. 1997a). This finding
has, however, been challenged by Vladilo (1998), who finds that the
(dust/metals)$_{DLAs}$=0.6(dust/metals)$_{MW}$ over the entire
metallicity range 0.1--1~Z$_{\odot}$. Vladilo's result highlights the
possibility that DLAs may not be representative of the metallicity
evolution of galaxies (see Pettini et al. 1998b). The issue is still
open, though, as current samples may select against low-redshift,
high-metallicity DLAs if the increasing dust content hampers detection
(Boisse` et al. 1998). The theoretical models of Dwek (1998) seem to
favor a modestly evolving dust/metals ratio, which goes from $\sim$1/2
to its equilibrium value within about 1~Gyr or so, depending on the
star formation history; the same timescale characterizes the slight
($\sim$50\%) overabundance of silicate dust over carbon dust when low
mass stars have not evolved yet. As a working hypothesis, our model
contains a quadratic relationship between the dust/gas ratio and the
metallicity, normalized to match the observed DLAs value at
Z$_{gas}$=1/15~Z$_{\odot}$, and the Galactic value dust/gas=0.0056 at
Z$_{gas}$=Z$_{\odot}$. The quadratic relation does not necessarily
reflect the actual dust enrichment of galaxies, but it is the simplest
possible formulation given the poor constraining the observations
currently provide. We will discuss later the consequences of relaxing
this assumption.

\subsubsection{The Dust Opacity}

The dust opacity of a galaxy is determined by the combination of three  
ingredients: 1. the dust column density; 2. the extinction curve; 3. the 
geometrical distribution of the dust in the galaxy. 

The simplest formulation for the dust column density E(B$-$V) is:
\begin{equation}
E(B-V)\propto N(H)\times F(Z)
\end{equation}
where N(H) is the gas column density, and F(Z) is the quadratic
function of the metallicity defined in section~2.1.4 for the dust/gas
ratio. F(Z) describes the fact that low metallicities imply
dust/metals ratios lower than the Galactic value (Pettini et
al. 1997a). The proportionality constant is the Galactic conversion
factor between N(H) and E(B$-$V) (E(B$-$V)=N(H)/5.9E+21, Bohlin,
Savage \& Drake 1978). The average gas column density N(H) is
proportional to the fraction of residual gas in the galaxy and to the
galaxy surface mass density. The fraction of residual gas is given by
the balance between star formation, gas recycling, and
outflows/inflows. The average surface mass density at z=0
(section~2.1.1), $\Sigma_{star+gas}(0)\simeq
100$~M$_{\odot}$~pc$^{-2}$, is a lower/upper limit to the surface mass
density at higher redshift, if outflows/inflows are present. In
sections~3 and 4, we will see that $\Sigma_{star+gas}$(z) is a
factor $\sim$1.7(2.0) higher(lower) at z=10 than at z=0 in case of
outflows(inflows).

In order to infer the behavior of the average surface mass density
$\Sigma_{star+gas}$(z) as a function of redshift, we use the
observational properties on the largest sample of high redshift
galaxies available to-date: the Lyman-break galaxies (cf. Steidel et
al. 1996). These z$\sim$3 galaxies have median 1500~\AA~ luminosity
L$_{1500}$=6.3$\times$10$^{40}$~erg~s$^{-1}$ within the half-light
radius (Steidel et al. 1996) and angular half-light radii around
0$^{\prime\prime}$.2--0$^{\prime\prime}$.3 (Giavalisco et al. 1996,
1998), the latter corresponding to physical half-light radii of
1--3~h$_{50}^{-1}$~kpc for a large range of values of q$_o$. The
bolometric surface brightness within the half-light radius is then
SB$\simeq$2.1$\times$10$^{9}$~L$_{\odot}$~kpc$^{-2}$, and the
dust attenuation corrected surface brightness is
SB$\simeq$6--12$\times$10$^{9}$~L$_{\odot}$~kpc$^{-2}$ (Pettini et al.
1998a, Calzetti 1997). If star formation has been constantly active
for 0.5--1~Gyr (cf. Pettini et al. 1998a), the mass-to-light ratio is
M/L=0.015--0.025, and the average stellar surface mass density of the
Lyman-break galaxies is in the range 90--300~M$_{\odot}$~pc$^{-2}$.
To this, we have to add the surface mass density of the
gas. Technically, at z=3 more than 90\% of the mass is still in gas,
but a lower fraction could be present in the Lyman-break galaxies if
they are going through a major star-formation event. We ascribe to the
z=3 galaxies the conservative figure of $\sim$50\%~ gas fraction,
implying a total surface mass density
180--600~M$_{\odot}$~pc$^{-2}$. The $\Sigma_{star+gas}$(z) of the
Lyman-break galaxies is thus a factor up to 3, 6, 10 higher than the
instantaneous value predicted at z=3 by outflows, closed-box, inflows,
respectively; it is also a few times higher than local value
$\Sigma_{star+gas}(0)\simeq 100$~M$_{\odot}$~pc$^{-2}$. In addition,
the observed half-light radii of the Lyman-break galaxies are on
average a factor $\sim$2 smaller than the half-light radii of local
galaxies, which range between 3 and 6~kpc. The evolution of the
surface mass density and of the half-light radius between z=3 and z=0
can be due to one of two reasons: (1) Lyman-break galaxies are
fragments of galaxies which will merge into larger units and are
undergoing intense star formation (Lowenthal et al. 1997); (2)
Lyman-break galaxies are massive galaxies (Steidel et al. 1996) and
stellar formation is proceeding ``inside-out''. In the latter case,
larger surface mass densities and smaller radii at high redshift are
explained by the fact that star formation is concentrated in the
central regions of the galaxies, where the density is
higher. Whichever the reason, we use the observational results to
model the evolution of the effective surface mass density as a
monotonically increasing function of redshift relative to the
instantaneous value implied by the presence of
outflows/closed-box/inflows at the same redshift. We have only two
points to tie the evolution to, one at zero redshift and the other at
z$\sim$3, and little handle on the driving physical parameters. In
first approximation, we assume that the evolution is driven by the
fraction f$_{gas}$ of residual gas in the galaxy, in the range
0.5$<$f$_{gas}<$1. Figure~1 shows our parametrization of the evolution
of the ratio of the effective to instantaneous surface mass densities
$\Sigma_{star+gas, eff}$(z)/$\Sigma_{star+gas, inst}$(z). For
instance, the outflow/closed-box/inflows models of section~3 give
$\Sigma_{star+gas, inst}$(3)=155/100/75~M$_{\odot}$~pc$^{-2}$ due to
the presence/absence of outflows/inflows of gas, implying 
$\Sigma_{star+gas, eff}$(3)=420/270/200~M$_{\odot}$~pc$^{-2}$. We
stress that the ``extra'' evolution of the surface mass density
implied by $\Sigma_{star+gas, eff}$(z)/$\Sigma_{star+gas, inst}$(z) is
purely observational.  Gas fractions f$_{gas}<$0.5 generally
correspond to redshift z$\le$1, and we assume that at this stage the
effective surface mass density is identical to its instantaneous
value. This assumption of no-evolution for z$\le$1 is supported by the
Medium Deep Survey galaxy sample of Phillips et al. (1995); the sample
has median redshift z$\simeq$0.3, and the galaxies show no evolution
of either the half-light radius or of the disk-to-total ratio.

The effective dust column density thus evolves as:
\begin{equation}
E(B-V)_{eff}\propto N(H) \times F(Z) \times
{\Sigma_{gas+star, eff}(z)\over \Sigma_{gas+star, inst}(z)}.
\end{equation}

As extinction curve, we adopt the Small Magellanic Cloud curve
(Bouchet et al. 1985), which is currently the only one known for a low
metallicity (Z$\simeq$0.1~Z$_{\odot}$), star-forming galaxy; we adopt
this curve as the best available for describing high redshift
galaxies. The choice of an extinction curve is important only for
$\lambda<$2800~\AA; at redder wavelengths, the three known mean
extinction curves (Galactic, LMC, and SMC, see, e.g., Fitzpatrick
1986) give similar attenuation values to within 5\%~ for a mixed geometry
(see below) and for E(B$-$V)$\le$2, well within the parameter space we
explore.

When dealing with galaxies as a whole, the simplest geometry is a
homogeneous mixture of dust, gas, and stars (Witt et al. 1992). This
has been shown to be valid in the low redshift regime for disk
galaxies (Xu \& Buat, 1995, Wang \& Heckman 1996), and for starburst
galaxies when the underlying old stellar population is considered
together with the young starburst population (Buat \& Burgarella
1998). Whether this is true in the high redshift regime, where all
stellar populations are young and the entire galaxy may be acting as a
single starburst region (Calzetti 1997a, Meurer et al. 1997), is not
clear. We will assume in first approximation the mixed geometry to be
valid at all redshifts. Among the solutions presented in section~3,
one case (Model~5) will have this assumption relaxed for high redshift
galaxies; other geometries will be discussed in
section~4.2. Observations of local galaxies indicate that the ratio of
the scale heights of stars and dust is a function of the
wavelength. In particular, the stars responsible for the UV emission
of a galaxy have the same scale height as the dust, h$_{star,
UV}$=h$_{dust}$ (e.g. Wang \& Heckman 1996), while the optical
emission from a galaxy is generally more vertically extended,
h$_{star, opt}\simeq$2--4~h$_{dust}$ (e.g., Kylafis \& Bahcall
1987). This is in agreement with the general result that dust is
mostly concentrated in the arms of spiral galaxies, where star
formation is occurring (White, Keel, \& Conselice 1997, Gonzalez et
al. 1998). We adopt two different functional behaviors for the ratio
h$_{star}$/h$_{dust}$ (Figure~2): one (hh1) which varies from 1 in the
UV to 2 at I, with a sharp change happening in the B band, and the
other (hh2) smoothly changing from 1 at 1,000~\AA~ to 4 at I. Both
functional behaviors are used throughout this paper. Finally, we are
modelling galaxies as flattened mass distributions and the final `net'
opacity (absorbed-to-total light ratio) is obtained from the average
of the model galaxy over all inclination angles.  The formalism for
the thin disk opacity model is detailed in Wang \& Heckman (1996). The
thin disk approximation does not take into account that nowadays about
1/3 of the galaxies are pure spheroids (ellipticals and S0s) and
contain about twice the mass of spirals (Persic \& Salucci 1992, see
also Schechter \& Dressler 1987 and Fukugita et al. 1998). We will
neglect this fraction because: (1) high-mass star formation nowadays
appears concentrated in flattened distributions (Scoville et al. 1998,
Downes \& Solomon 1998) and we are mostly interested in the opacity
suffered by the UV emission from massive stars; (2) the fraction of
spheroids may not be constant in time; Kauffmann, Charlot \& White
(1996) suggest that only 1/3 of the present-day spheroids are
assembled at z$\sim$1. If galaxies were formed by hierarchical merging
of substructures, this may indeed have occurred late in the life of
the Universe, at z$\le$1, and most of the high-redshift star formation
is occuring in disk-like structures (Kauffman 1996, Baugh, Cole \&
Frenk 1996). The presence of spheroids, in addition, has little
effect on the final average over the inclination angles: we estimate
that for h$_{star}$/h$_{dust}$=1 and E(B$-$V)$\le$1 the `net' opacity
decreases by a factor $\sim$1.25 if we allow for a mixture of 2/3
flattened disks and 1/3 spheroids.

\subsubsection{The Dust Emission}

The fraction of stellar light absorbed by dust at UV and optical
wavelengths is re-emitted in the FIR. In order to calculate the
intensity and spectrum of the dust FIR emission, we need to assume a
spectral energy distribution for the galaxy's stellar population and a
model for the dust emission.

We adopt for the stellar population a 1~Gyr old continuous star
formation SED with solar metallicity (Bruzual \& Charlot 1996).  The
dependence of the FIR emission on the age of the stellar population is
negligible for the constant star formation case, because the FIR flux
receives the bulk of the contribution from the UV emission of the
stars. However, the 1~Gyr population will generally be too blue at
optical wavelengths in the latest stages of the Universe, as in our
cosmology the Universe today is 13.3~Gyr old.  Thus, for sake of
completeness, we will also show in section 4.2  the case of a 5~Gyr
continuous star formation population. Whichever the model adopted for
the galaxy SED, the UV spectrum is normalized to the observed
luminosities. The normalization is from the observed flux density at
2,800 \AA~ in the range z=0--1 (Lilly et al. 1996), after the model
SED is convolved with the opacity at the same wavelength. The fraction
of ionizing photons absorbed by dust is given by the formula reported
in Calzetti et al. (1995).  Despite the low metallicity SEDs being
generally bluer than the high metallicity SEDs, the decreasing 
average metallicity at higher redshift has a relatively minor
impact on the average galaxy SED, on the basis of current population
synthesis models. The difference in the number of ionizing photons for
metallicities between 0.1 and twice Z$_{\odot}$ is about 13\%, and the
difference in the normalized UV flux is about 11\% (Leitherer \&
Heckman 1995).

The dust emission spectrum is by no mean well known or easily
parametrized in local galaxies. Our Milky Way shows a complex
combination of blackbody temperatures (Cox, Kr\"ugel \& Mezger 1986,
D\'esert, Boulanger \& Puget 1990), and other galaxies as well are
better described by multiple temperature components, with the most
simplified models using at least two (Lonsdale-Persson \& Helou 1987,
see also the ISO data of Krugel et al. 1998). For the purpose of this
study we consider the simplest possible case: a single-temperature
blackbody distribution convolved with a dust emissivity $\propto
\nu^{\epsilon}$, where $\epsilon=2$, bearing in mind that this is an
oversimplification. The temperature of the dominant dust emission
component is correlated with the star formation activity in low
redshift galaxies (Helou 1986); in starburst galaxies, the observed
correlation is between the temperature and the FIR surface brightness
of the star forming region (Lehnert \& Heckman 1996), implying a
correlation between temperature and the surface density of the star
formation. We reproduce this trend for the dust temperature with:
\begin{equation}
T(z)\propto \biggl(SFR_{int}(z)\,{\Sigma_{star+gas, eff}(z)\over
\Sigma_{star+gas, inst}(z)}\biggr)^{1/(4+\epsilon)},
\end{equation}
where T(z) is the blackbody temperature as a function of redshift and 
SFR$_{int}$(z) is the intrinsic star formation rate;
$\Sigma_{star+gas, eff}$(z)/$\Sigma_{star+gas, inst}$(z) measures 
how ``concentrated'' the star formation is and has value 
$\approx$3 at z=3 (see section~2.1.5 and
Figure~1). SFR$_{int}$(z)\,$\Sigma_{star+gas,
eff}$(z)/$\Sigma_{star+gas, inst}$(z) is a measure of the surface star
formation density, and, thus, of the surface brightness of the star
forming region.  The proportionality constant for Equation~5 has been
chosen to recover the mean dust temperature of local quiescent
galaxies (T$\sim$20~K).

\subsection{Outputs from the Model}

The three main outputs from this model of galaxy evolution are:
\begin{enumerate}
\item the intrinsic SFR density as a function of redshift; this is the 
result of the iterative process to yield consistent solutions 
between the dust opacities and the correction factors to the 
UV-derived global SFR density; 
\item the dust opacities, defined as the fraction of emerging to
total stellar light, at UV and blue wavelengths (galaxy restframe) as
a function of redshift;
\item and the spectral energy distribution of the CIB in the range
100--2,000$\mu$m (observer's restframe), from the integral along the
line of sight of the dust emission from galaxies.
\end{enumerate}
These outputs are compared with available data. The strongest
constraints come from observations of the Local UV opacity (e.g., Wang
\& Heckman 1996), from the H$\alpha$- and ISO-derived intrinsic SFR
density at low redshift (Gronwall 1998, Hammer \& Flores 1998), and
from the measured CIB from COBE DIRBE and FIRAS (Hauser et al. 1998,
Fixsen et al. 1998). Predictions from intermediate steps of the model,
namely the evolution of the metal content, of the gas column density,
and of dust temperature, and the local FIR energy density are also
compared with available observational data.

\section{Results}

Galaxy evolution is first followed by using the observed SFR(z) as
star formation history (Model~1, see Figure~3). The main properties of
the model are listed in Table~1. The observed SFR(z) is unable to
reproduce the basic characteristics of the local Universe: the amount
of stars produced by z=0 is $\Omega_{star}=0.0012$~h$_{50}^{-1}$, or
24\%~ of the total stellar mass observed today in galaxies. The
residual gas fraction at z=0 is 80\%, well above the local value of
15\%, and, as consequence, the final metallicity of the gas is about
1/6~Z$_{\odot}$. The final metallicity of Model~1 is so much below
solar that outflows/inflows are not required to regulate it, and
galaxies have been treated as closed boxes. To create a reference
baseline, the surface mass density has been assumed non evolving,
$\Sigma_{gas+star, eff}$(z)/$\Sigma_{gas+star, inst}$(z)=1. From the
simple act of forming stars, Model~1 predicts non-zero opacities. In
particular the emerging 2800~\AA~ and 1500~\AA~ fluxes from galaxies
at z$\le$0.1 are about 1/2 and 1/3 of the total, respectively
(Figure~4). Thus, despite the small final metallicity, the low
redshift opacities are relatively large; this is due to the large
residual gas fraction. The presence of non-zero UV opacities is, of
course, in contradiction with the assumption that the observed,
UV-derived SFR(z) is equivalent to the intrinsic SFR(z), namely
SFR$_{obs}$/SFR$_{int}$=1. The magnitude of the contradiction is shown
in Figure~4 itself, where SFR$_{obs}$/SFR$_{int}$ is compared with the
predicted emerging-to-total light at the appropriate wavelength.  The
FIR background implied by those opacities (Figure~5) is $\le$25\%~ of
the observed values from COBE DIRBE$+$FIRAS results. This comparison,
although necessary for sake of clarity, is artificial, because by
assumption Model~1 does not contain dust, and therefore there cannot
be a CIB. As a general consideration, Model~1 produces far too little
mass permanently locked in stars; even an IMF with a Salpeter slope
down to 0.1~M$_{\odot}$ would only produce a factor 1.64 more mass in
stars. Thus the star formation rates must be higher than directly
measured.

The predicted UV opacities of Model~1 provide guidelines for the
iterative procedure to produce self-consistent corrections to
SFR$_{obs}$, the observed SFR(z).  The solution the procedure
converges to is not unique; with our limited knowledge of SFR$_{obs}$,
which consists of 6 observational data points (i.e., excluding the z=4
one), we find two solutions (Model~2 and Model~3) to SFR$_{int}$, when
we allow for either closed-box or outflow models. The case of inflows
is also discussed (Model~4), but less extensively, because the
solution is somewhat intermediate between Model~2 and Model~3. In a
fourth solution (Model~5), we impose the SFR$_{obs}$ point at z=4 to
be a constraint, and we also allow for variation of some of the input
parameters.  Model~2 and Model~3 bracket two extreme behaviors of the
intrinsic SFR(z), for the selected input parameters. However, as we
will see in the next section, one of the two satisfies the available
observational constraints better than the other, within the limits of
our assumptions. All Models allow the effective surface mass density
of galaxies to evolve with the residual gas fraction f$_{gas}$ (see
Figure~1). The main characteristics of the four solutions are
summarized in Table~1.

The SFR$_{int}$ of Model~2 (Figure~3) dictates that star formation
increases from z=10 (our fiducial starting point for star formation)
to z$\simeq$1.2, and then decreases from z$\simeq$1.2 to the
present. The global shape is similar to SFR$_{obs}$, but the
correction factors are different at high and low redshift. In
particular, at z=2.75 SFR$_{int}$=1.8$\times$SFR$_{obs}$, while at
z$<$1 SFR$_{int}$=3.9$\times$SFR$_{obs}$; galaxies are, therefore,
more opaque at recent times than at high redshift. As mentioned
earlier we do not attempt to use the SFR$_{obs}$ at z=4 as a
constraint either in this Model or in Models~3 and 4.  The final
metallicity of Model~2 is 3/4~Z$_{\odot}$, implying that excess metals
are not produced; thus galaxies are treated as closed boxes. The
residual gas fraction at z=0 is f$_{gas}$=0.34; this value is larger
than the target of 15\%, but we consider it marginally acceptable in
view of the large uncertainties affecting each of our input
parameters. The mass in stars is 78\% of the target value. The dust
column density is an increasing function of time and a similar
behavior is displayed by the opacities (Figure~6). The predicted
2800~\AA~ emergent-to-total radiation in the local Universe is around
a factor 0.24 (Figure~7).  Such high dust opacity at low redshift
implies comparably low redshifts for the main contributors to the CIB
radiation; Figure~6 shows the contribution to the flux of the CIB as a
function of redshift: most of the contribution to the observer's
restframe emission at 140~$\mu$m comes from galaxies at z$\sim$1,
while the contributors to the 240~$\mu$m emission are more smoothly
distributed with median redshift z$\sim$1.4. At longer wavelengths,
the flux contribution to the CIB is smoothly distributed at redshift
z$>$1, and is almost constant at 850~$\mu$m. The dust temperature
peaks at z$\simeq$1.2, following the shape of SFR$_{int}$ (see
Equation~5). Model~2 agrees with observations of the CIB at short
wavelengths ($\lambda<$400~$\mu$m), but falls short of the data at
longer wavelengths (Figure~8); the predicted FIR SED is thus too hot
in comparison with data, in agreement with the hottest galaxies being
located at relatively small redshift (Figure~6).

The second solution which satisfies the self-consistency requirement
between SFR$_{obs}$/SFR$_{int}$ and the UV opacities is shown in
Figures~9 to 11 (Model~3). The SFR$_{int}$ of Model~3 is completely
different from Model~2 (Figure~3): it is constant at high redshift and
decreases below z$\simeq$1.2, roughly resembling the SFR(z) of
Rowan-Robinson et al. (1997). In numbers: at z=2.75
SFR$_{int}$=5.8$\times$SFR$_{obs}$ and at z$\le$1.2
SFR$_{int}$=3.2$\times$SFR$_{obs}$. In Model~3, the residual gas
fraction at z=0 is 13\%, practically meeting the target of 15\%. The
large amount of star formation implies that outflows are invoked to
maintain the final gas metallicity in galaxies to solar values; thus
galaxies were on average 1.7 times heavier at z=10 than at z=0
(cf. Edmunds \& Phillips 1997), and $\Omega_{gas}$=0.0082 at z=3. The
resulting rotation velocities at z=3 are $\sim$250~km~s$^{-1}$, for a
constant mass in dark matter and for a local typical rotation speed of
210~km~s$^{-1}$, thus still perfectly reasonable. In the absence of
outflows, the final metallicity of the galactic gas would be
Z=1.44~Z$_{\odot}$. The dust column density peaks at z$\simeq$1.8
(Table~1 and Figure~9), due to the decreasing gas fraction despite the
still increasing metallicity of the galaxy; the ratio of the
emerging-to-total light from galaxies follows an analogous trend. This
trend is in agreement with the correction factors applied to
SFR$_{int}$ (Figure~10), as expected from the fact that Model~3 is a
solution to our iterative procedure. The large opacities at high
redshift imply that the contributors to the CIB in Model~3 are spread
over a larger redshift range than in Model~2 (Figure~9). The peak of
the contribution at 140~$\mu$m is around z=1.2, not very different
from Model~2, but the contribution to the 240~$\mu$m CIB emission is
spread over the entire redshift range sampled, with a median at
z$\simeq$2.4. Even more extreme is the redshift dependence of the flux
contributions at 450~$\mu$m and 850~$\mu$m, which increase with z. The
predicted FIR background is given in Figure~11, where it is compared
with the observed CIB results. The agreement between
the CIB data and the prediction of Model~3 is much better than in
Model~2, and is within the 1~$\sigma$ error bar of the observations
for the entire spectral range up to 1600~$\mu$m. The Model~3 spectrum
still tends to be slightly hotter than the observational points at the
longest wavelengths.  However, the long wavelength regime is also the
region where the FIRAS data have the largest uncertainties, thus it is
hard to estimate the real disagreement between model and data.

When inflows of metal-free gas, rather than outflows of metal-rich
gas, are invoked to constrain the metallicity of the galactic gas to
solar values at z=0, the iterative procedure converges to a solution
(Model~4) intermediate between Model~2 and Model~3 in the low-z regime
and close to Model~3 in the high-z regime. In Model~4,
SFR$_{int}$=3.5$\times$SFR$_{obs}$ at z$<$1, and
SFR$_{int}$=4.4$\times$SFR$_{obs}$ at z=2.75 (Figure~3). The final gas
fraction here is f$_{gas}$=0.19, very close to the target value;
without inflows the final metallicity would be Z=1.17~Z$_{\odot}$,
slightly above solar. The initial mass in galaxies at z=10 is a factor
0.68 of the local value, and $\Omega_{gas}$=0.0035 at z=3.  The dust
column density peaks at z=0.4 (Figure~12), and the opacities follow
the same trend; the result of the iterative procedure can be seen in
Figure~13, where the ratio SFR$_{obs}$/SFR$_{int}$ is shown to match
the emerging-to-total UV light. Model~4 reproduces the CIB
observations with a similar level of accuracy as Model~3 (Figure~14),
thanks to the large UV opacities at high-z.

For the solution Model~5 (see Figure~3 and Table~1), the value of
SFR$_{obs}$ at z=4 is used as an additional constraint.  In this
model, we also explore the effect of changing some of the input
parameters. The true yield $\alpha_Z$ is increased from 0.7 to 1.0,
the latter being possibly closer to the actual yield of stars. The
stellar populations in high-z (z$\ge$2.75) galaxies are completely
embedded in, rather than uniformly mixed with, dust, possibly a better
representation of primeval galaxies. Finally, we assume that the
dust/metal ratio does not evolve, but is a constant equal to the Milky
Way value. The effect of these assumptions is shown in Figure~15: the
peak of the dust column density is much higher, by a factor $\ge$3,
than in the other three models, clearly reflecting the effect of the
larger amount of metals produced (higher $\alpha_Z$) and of the larger
fraction of dust created (dust/metals=constant). The match between the
SFR$_{obs}$/SFR$_{int}$ ratio and the emergent-to-total UV light
(Figure~16) is achieved by producing 111\% of the target mass in stars
and leaving only 6\%~ of residual gas. Even so, the agreement between
SFR$_{obs}$/SFR$_{int}$ and the emergent-to-total light is not as good
as in the three previous cases: a better match would require an
increase of $\sim$10\% of the SFR$_{int}$ at z$<$1.2; this is however
problematic because the model would immediately run out of gas. The
large amount of metals produced requires large outflows to constrain
the final metallicity of the galaxy to Z$_{\odot}$: the initial
baryonic mass in galaxies needs to be 4 times larger than at z=0. In
the absence of outflows the final gas metallicity would be
Z=2.90~Z$_{\odot}$. Model~5 does not reproduce the spectral energy
distribution of the CIB as well as Models~3 or 4 (Figure~17).
However, the discrepancy between predictions and data in Figure~17
disappears almost completely if we change the dust emissivity index
from $\epsilon$=2 to $\epsilon$=1.5.

The mixed dust/star geometry adopted in Models~2--4 and in the low
redshift regime of Model~5 produces a grey effective extinction,
namely, a mild reddening coupled with large global attenuations. This
is a general characteristic of mixed geometries (Witt et al. 1992);
its effect is shown in Figure~18, where the attenuation at the peak
column density E(B$-$V)$_p$ for each of the four Models (Table~1) is
compared with the extinction produced by a foreground screen of dust
with an SMC curve and a modest E(B$-$V)=0.1. The reddening between
1000~\AA~ and 10000~\AA~ produced by E(B$-$V)$_p\sim$0.3 in
Models~2--4 is comparable with the case of a foreground screen with
E(B$-$V)=0.1, but the global attenuations are a factor $\sim$1.5
larger. As a result, the galaxies in our model Universe will appear
blue, even in the presence of non negligible dust obscuration. Model~5
stands out from the general trend of the other three models because
its input parameters dictate larger global dust column densities.

In the next section, we will concentrate the discussion on the
differences between Model~2 and Model~3, which represent two extreme
behaviors. We recall Model~4 only when needed for completeness, as the
general behavior of the latter is relatively close to Model~3. The
same will be done for Model~5, although this solution implies rather
extreme conditions at z$<$1, in the form of large opacities.

\section{Discussion}

\subsection{The Intrinsic Star Formation History}

Models~2 and 3 imply drastically different star formation histories
for the Universe, although they are both reasonable solutions for our
model of galaxy opacity evolution (Figures~7 and 10). In Model~2,
34\%~ of the star formation happens at redshift z$\ge$1.2, and only
2\% of the stars are produced at z$>$3. In Model~3, 57\%~ of the star
formation happens at redshift z$\ge$1.2, with 20\% of the stars
produced at z$>$3. The difference between the two models is even more
obvious when redshifts are converted in time, since Model~2 and
Model~3 imply that 1/3 and more than 1/2, respectively, of the star
formation happened during the first $\sim$4~Gyr since the beginning of
the Universe (for our adopted cosmology). For comparison, Model~5
produces 1.5\% of the total mass in stars at z$>$3 and 45\% at
z$\ge$1.2.

Model~2, however, reproduces the CIB spectral energy distribution less
accurately than Model~3 (cf. Figure~8 with Figure~11). Model~2 would
not necessarily be in better agreement with the CIB if a value of the
dust emissivity $\epsilon<$2 were adopted, especially if we consider a
more realistic model of the galaxies FIR emission. Local galaxies are
characterized by a broad distribution of dust temperatures, of which
our assumed T=20~K is most likely towards the low end of the
range. Local galaxies have a FIR distribution skewed towards higher
energies and spread over a larger range of frequencies than assumed
here, with peaks around wavelengths $\le$140~$\mu$m (e.g., Krugel et
al. 1998). Even if $\epsilon$ is less than 2 in actual galaxies, this
does not guarantee that the global emission will have more power at
long wavelengths, because of the broad FIR energy distribution. Most
likely, Model~2 will still be unable to reproduce the CIB at long
wavelengths.

As discussed in the previous section, Models~2 and 3 predict different
redshift dependences of the flux contribution to the CIB at long
wavelengths. For instance, Model~2 gives a rather flat 850~$\mu$m
contribution for z$>$1 (cf. Figure 6), while in Model~3 the
contribution is an increasing function of redshift. More in detail,
$<$Flux(850, 2$<$z$\le$5)$>$=3.55\,$<$Flux(850 0.9$\le$z$\le$2)$>$
in Model~3 and $<$Flux(850, 2$<$z$\le$5)$>$=0.92\,$<$Flux(850,
0.9$\le$z$\le$2)$>$ in Model~2. The redshift range 2$\le$z$\le$5 spans
a time interval of 1.7~Gyr, while 0.9$\le$z$\le$2 spans an interval of
2.5~Gyr. Since we assume that the number density of galaxies is a
constant, such differences in background flux are due to variations in
the total FIR luminosity of the sources in the band. In Model~3, where
the SFR density is constant for z$>$1, the flux contribution to the the
850~$\mu$m band is determined by the slight increase of the dust
temperature with redshift and the k-correction within the band. For
Model~2, the contribution to the 850~$\mu$m is a combination of
k-corrections and both variable SFR$_{int}$(z) and T(z).  The SCUBA
sources, which sample the bright end of the FIR luminosity function,
should be mostly located at z$>$2 if Model~3 is correct, or be roughly
homogeneously distributed as a function of redshift if Model~2 is
correct. The current (uncertain) redshift placements give 3--5 sources
at 0.9$\le$z$\le$2 and 8--10 sources at z$>$2 (depending on where we
place the two sources of Barger et al. 1998), implying a ratio
1.6--3.3 with a median value of $\sim$2.3 between the two redshift
bins. The result is still inconclusive, and must await further
constraints on the redshifts of the sources. For reference, the
reddest (dustiest) z$\sim$3 Lyman-break galaxies in the Hubble Deep
Field are below the detection limits of the SCUBA observations; 
their predicted fluxes are below 1.7~mJy and mostly around $\sim$1~mJy
(Meurer et al. 1998), while the SCUBA-HDF field has a 5~$\sigma$
detection limit of 2~mJy (Hughes et al. 1998).

As mentioned in Section~2.1.3, the SFR$_{obs}$ data point at z$\sim$4
has not been used as a constraint for Models~2--4 due to the potential
underestimate introduced by incompleteness and volume corrections.  A
posteriori, we can infer that Model~2 implies that the observed
z$\sim$4 point should be underestimated by a factor $\sim$1.5, while
Models~3 and 4 imply that this data point is underestimated by a
factor $\sim$3.

We now compare the predicted SFR$_{int}$ for the local Universe
(Models~2 to 5) with the results of observational surveys in H$\alpha$
or radio, which are less sensitive to dust obscuration than the UV
emission. All our solutions give the local SFR$_{int}$ between a
factor 3 and 4 of the value measured from UV observations, namely
SFR$_{int}$=0.0095-0.0125M$_{\odot}$~yr$^{-1}$~Mpc$^{-3}$ at a median
redshift of 0.05, with the lowest value given by Model~3.  The local
($<z>$=0.05), reddening-corrected H$\alpha$ luminosity density of
Gronwall (1998) correponds to an intrinsic
SFR=4.1$\times$10$^{-42}$~L(H$\alpha$)$\sim$0.008~M$_{\odot}$~yr$^{-1}$~Mpc$^{-3}$
for our IMF; the analogous data point from Gallego et al. (1995) gives
SFR$\sim$0.005~M$_{\odot}$~yr$^{-1}$~Mpc$^{-3}$. The data points carry
a typical uncertainty of $\sim$50\%~ (e.g., Gallego et
al. 1995). Radio observations of local galaxies trace the supernova
rate and, indirectly, the star formation rate at a wavelength which is
not affected by dust obscuration; the SFR density inferred from observations
at 1.4~GHz is $\sim$0.016~M$_{\odot}$~yr$^{-1}$~Mpc$^{-3}$ on our IMF
scale, again with $\sim$50\%~ uncertainty (Cram 1998). Therefore, our
SFR$_{int}$ is in reasonable agreement with the data points from both
H$\alpha$ and radio observations.

Another important check is whether the amount of dust implied by our
solutions overpredicts the FIR emission from local galaxies, the
latter being measured by IRAS. The FIR emissivity at 60~$\mu$m in the
local ($<z>$=0.01) Universe is
$\nu$F($\nu$)$\simeq$3.4$\times$10$^7$~L$_{\odot}$~Mpc$^{-3}$
(Villumsen \& Strauss 1987, Soifer \& Neugebauer 1991). The total FIR
energy density in the wavelength range 8--1000~$\mu$m from our
Models~2--4 is
E(8--1000~$\mu$m)$\simeq$6.9--9.3$\times$10$^7$~L$_{\odot}$~Mpc$^{-3}$
at the same median redshift.  Using a conversion factor of 2.67
between total energy and power at 60~$\mu$m (cf. Soifer \& Neugebauer
1991), the predicted total FIR emissivities in the models correspond
to predicted 60~$\mu$m emissivities of
$\sim$2.6--3.5$\times$10$^7$~L$_{\odot}$~Mpc$^{-3}$, which bracket the
observations.

Infrared observations also offer an independent measure of the
intrinsic SFR density of the Universe. ISO observations at 6.75~$\mu$m
and 15~$\mu$m of one of the CFRS fields have been recently used to
derive preliminary values of the SFR over the redshift range
0.3$\le$z$\le$1 (Hammer \& Flores 1998). The SFR$_{int}$ of Models~2
and 3 are a factor 1.7$\pm$0.25 and 1.4$\pm$0.2, respectively, larger
than the ISO values in the same redshift range. The difficulty with
this type of observations is in the uncertain derivation of a SFR from
the restframe $\sim$10~$\mu$m emission of galaxies. However, if
confirmed by further data, the excess SFR$_{int}$ at low-z given by
our solutions may imply that either our galaxy model is too simplistic
in the low redshift regime, where the dust is mixed with multiple-age
populations; or that the intrinsic SFR at high redshift is even higher
than what required by Model~3. 

Relatively low UV opacities at z$<$1 require, indeed, a high gas
consumption rate at high redshift, namely high SFR$_{int}$. The
average dust column density of a galaxy, E(B$-$V), is the result of
two opposite effects, both monotonically dependent on time: increase in
metallicity, which increases E(B$-$V), and decrease of the gas
fraction, which decreases E(B$-$V). The peak at
z$\simeq$1.8 for the dust column density in Model~3 is a consequence
of those opposite processes. Low opacities at z$<$1 thus require that
a large fraction of the gas has been already locked in stars. For this
to happen, SFR$_{int}$ at high redshift must be large, which also
induces a faster metal enrichment of the galaxy's ISM, like in
Model~3. As a result, high-z galaxies will tend to be opaque. The UV
attenuation A$_{1500}=$1.85~mag predicted by Models~3 and 5 at z=3 is
in agreement with the recent results by Pettini et al. (1998a), who
find attenuation values A$_{1500}\simeq$1--2~mag in six Lyman-break
galaxies, from joint measurements of rest-frame UV spectra and optical
nebular line emission.

Various authors (e.g. Dickinson 1998, Meurer et al. 1997, Calzetti
1997a, Pettini et al. 1998a) have attempted to correct the UV emission
of the Lyman-break galaxies at z=3 using the `starburst' reddening
curve derived by Calzetti et al. (1994). In particular, the UV spectra
of these galaxies appear reddened by a color excess E(B$-$V)$\sim$0.15
if the functional shape of Calzetti (1997a) is used. We have plotted
in Figure~19 the UV attenuation A($\lambda$) estimated for the
Lyman-break galaxies using the `starburst' reddening curve and
E(B$-$V)=0.15 and, for comparison, the effective UV attenuation
predicted by Model~3 at z=3. The two curves have a similar shape and
the difference between the two attenuations is less than 13\%~ over
the entire wavelength range 1200-3000~\AA. The implication is that,
for all practical purposes, the two curves are equivalent.

The dependence of the dust temperature on SFR(z) (Equation~5) causes
the dust to be hotter in high-redshift galaxies than in low-z galaxies
for Models~3 and 4 (e.g., Figure~9); however, the functional
dependence is very weak, as it goes as a power 1/(4$+\epsilon$), so
that the hottest galaxies have dust temperature between 35 and 40~K in
all models. This is the typical dust temperature of local moderate
starburst galaxies, like e.g. NGC7673 (Calzetti et al. 1995, Helou
1986).  This clumpy irregular has a star formation rate of
$\sim$4~M$_{\odot}$~yr$^{-1}$ within its central $\sim$2~kpc (Calzetti
1997b). For comparison, Lyman-break galaxies at z=3 have an average
observed SFR(1,500~\AA)=5~M$_{\odot}$~yr$^{-1}$ within their
half-light radii (Steidel et al. 1996), which corresponds to an
attenuation-corrected SFR(1,500~\AA)=28~M$_{\odot}$~yr$^{-1}$ for
Model~3, or a star formation density of
1~M$_{\odot}$~yr$^{-1}$~kpc$^{-2}$, similar to that found in NGC7673.

The measured metallicities of DLAs are $\sim$1/15~Z$_{\odot}$ in the
redshift range 2$<$z$<$3 (Pettini et al. 1997b) and
$\sim$1/5--1/10~Z$_{\odot}$ at 0.5$<$z$<$1 (Boiss\`e et al. 1998, Pettini
et al. 1998b). As pointed out by Boiss\`e et al., the measurements at
z$\sim$1 could be a lower limit to the actual abundances of DLAs at
those redshifts because dust obscuration will bias observations
preferentially towards metal-poor systems. Assuming that the gas
abundances in DLAs are representative of galaxies (see, however,
Vladilo 1998 and Pettini et al. 1998b), Models~2, 3, 4 and 5 predict
that the metallicity of the gas is $\sim$1/40, 1/10, 1/6,
1/60~Z$_{\odot}$, respectively, in the redshift range 2$<$z$<$3 and
$\sim$0.3, 0.4, 0.6, 0.25~Z$_{\odot}$, respectively, at
0.5$<$z$<$1. Among the four solutions, Model~3 gives the closest
metallicity values to the observational data at high
redshift. However, metal enrichment depends on the highly uncertain
value of the effective yield $\alpha_Z$, which, in turn, depends on
the stellar IMF (see discussion in Pagel 1987, 1998, Kauffmann \&
Charlot 1998). Thus the underabundance of Model~2 at high redshift may
simply be a consequence of the uncertainty in the metal yield. The gas
column densities at z=3 implied by our solutions 2, 3, and 4 are in
the range N(H)=(5.4--12)$\times$10$^{21}$~cm$^{-2}$, with the lowest
value corresponding to Model~4 and the highest to Model~3. This range
is comparable to the observed one,
N$_*$(HI)=(2--10)$\times$10$^{21}$~cm$^{-2}$ (Storrie-Lombardi,
McMahon \& Irwin 1996), once we take into account that our model gives
the total column density in gas, while observational measurements of
HI do not include molecular hydrogen. The z=3 gas column density of
Model~5 is N(H)=3.5$\times$10$^{22}$~cm$^{-2}$, much higher
than what measured; this large value is a consequence of the large gas
outflows required in Model~5 to constrain the final gas metallicity to
Z$_{\odot}$. Gas inflows/outflows imply
$\Omega_{gas}$(z=3)$\sim$0.0035--0.008 in Models~4 and 3,
respectively. The gas density inferred from DLAs at the same redshift
is $\Omega_{gas}$(z=3)$\sim$0.0055, after correction for some
reddening and for 20\% gas fraction not in DLAs (Storrie-Lombardi et
al. 1996). The measured value is a factor 1.5 higher/lower than
predicted by our models. Allowances for uncertainties and/or
incompletenesses in the observations may in principle reduce some of
the discrepancy. Another way to circumvent this difficulty is to allow
in the models contemporary presence of inflows of metal-free gas in
conjunction with outflows of metal-enriched gas. The outflows of
Model~3 pollute the IGM with metals to the modest value of
Z$_{IGM}$=0.05--0.07~Z$_{\odot}$, while Model~5 gives
Z$_{IGM}$=0.08--0.11~Z$_{\odot}$, in the redshift range
0.5$\le$z$\le$0. These values are a factor $\sim$3--4 smaller than what
found in the intracluster medium (e.g., Mushotzky \& Loewenstein 1997,
Renzini 1997), but are comparable with observations of the Ly$\alpha$
Forest at z$\sim$0.5 (Barlow \& Tytler 1998). Again, Z$_{IGM}$ depends
on the highly uncertain value of $\alpha_Z$ (see below).

The shape of SFR$_{int}$ of Model~3 resembles the `monolithic
collapse' formation model for stellar populations. About 20\%~ of the
total mass in stars today is produced at z$>$3 (within 2~Gyr from the
beginning of the Universe). If $\sim$60\%~ of the local stellar mass
is in old spheroids (Persic \& Salucci 1992, Schecter \& Dressler
1987), then 33\%~ of the stars which are today in spheroids were
formed at z$>$3. These values agree with Renzini (1998), who argues on
the basis of the intracluster medium metallicity that about
30\%~ of all stars were produced at z$>$3.

The functional shape of Model~3 is different from what found for,
e.g., radio-loud (flat-spectrum) quasars (Shaver et al. 1996, 1998,
Hook et al. 1998). Radio observations of these objects indicate a peak
in space density around z=2.5; the decrease in density beyond z=2.5
has been suggested to be a real evolutionary effect (rather than dust
obscuration, as suggested by Webster et al. 1995). The evolution of
the space density of quasars resembles our Model~5. Shaver et al. have
likened the evolution of the radio-loud quasars' space density to the
galaxy star formation history. However, such connection is not
straightforward, as we don't know, for instance, the details of the
process which switches on a quasar, nor have we unraveled the relation
between AGN activity and galaxy evolution. It is not clear yet that
radio-selected samples are representative of the entire QSO
population; thus, the possibility that dust obscuration is responsible
for the high-z decrease in the space density of optically-selected
QSOs is not excluded yet.

\subsection{The Impact of Parameter Choices}

Using Model~3 as a baseline, we modify one by one some of the input
parameters in the model, to assess their impact on the results.

\subsubsection{The Evolution of the Dust/Metal Ratio}

Our model uses the observational results of Pettini et al. (1997a) to
infer the evolution of the dust/metals ratio with metallicity. If the
dust/metals ratio instead depends only on the physics of the dust
(e.g, the results of Vladilo 1998), and is constant and equal to the
Milky Way value, the consequences for our model are pretty
straighforward: low-metallicity systems will be more dusty. In
particular, the predicted emergent-to-total light ratios decreases at
all redshifts by $\sim$60--70\%. Thus, the effect of an evolving
dust/metals ratio is dramatic in low-metallicity objects.

\subsubsection{The Galaxy Spectral Energy Distribution}

Changing the input galaxy SED from a 1~Gyr constant star formation
model to a 5~Gyr constant star formation model increases the fraction
of optical-NIR bright stars relative to the UV. The CIB is the only
output which uses the galaxy SED in our model, and, therefore, is the
only quantity potentially affected. As mentioned in section~2.1.6, the
bulk of the contribution to the galaxy FIR emission come from the UV
stellar light reprocessed by dust; the contribution of the optical-IR
stellar radiation is small in comparison. Indeed, when the CIB is
computed from Model~3 using a constant star formation galaxy SED 5~Gyr
old, instead of 1~Gyr old, the difference between the results is
minimal: the intensity at $\lambda<$300~$\mu$m is increased by
10--15\%, while it is practically unchanged at longer wavelengths.

\subsubsection{The Dust Model}

The three basic assumptions on the dust model we have implemented are:
the dust is homogeneously mixed with the stars with a
wavelength-dependent scale height; the dust temperature depends on the
surface density of the star formation rate (Equation~5) ; the value of
the emissivity is $\epsilon$=2. These assumption are relaxed in turn
in this section.

The homogeneous mixture of dust and stars appears to be a good
representation of local galaxies (Xu \& Buat 1995, Wang \& Heckman
1996). However, high redshift galaxies may obey other geometrical
descriptions; for instance the recently formed stars may be surrounded
by, rather than mixed with, dust. This case has been implemented above
in Model~5. Here, in order to bracket a range of situations, we
consider two different geometries: 1. the dust surrounding the stellar
population is in a uniform shell; 2. the dust surrounding the stellar
population is distributed in clumps with a covering factor
$\sim$50\%. The results are shown in Figure~20.  The UV opacities are
overpredicted by the model of uniform dust (u) and underpredicted by
the clumpy dust (c). In this sense, the two models bracket a range of
dust geometries for galaxies, expecially for the high redshift ones.
It should be noted that the mixed geometry adopted throughout this
paper produces opacities which are roughly the `middle point' between
the two geometries discussed in this section.

Observations suggest that the `average' temperature of the dust in a
galaxy is correlated with the density of the star formation activity
(Lehnert \& Heckman 1996). This behavior is reproduced in our model
via Equation~5. The correlation is relaxed here to investigate its
impact on the results. Figure~21 shows the case in which we adopt a
dust temperature T=21~K at all redshifts and independent of the star
formation activity. A `cool' FIR emission from high redshift galaxies
in Model~3 implies too much flux in the CIB at long wavelengths
($\lambda>$250~$\mu$m) and too little flux at short wavelengths. This
is understandable as the high redshift galaxies contribute to the
longest wavelength part of the CIB.

Finally, we change the dust emissivity from $\epsilon$=2 to
$\epsilon$=1. These two values bracket the range observed in Galactic
dust; the exact value of the dust emissivity is still unknown,
although some authors favor values around $\epsilon$=(1.5,2.0) at long
wavelengths (e.g., D\'esert et al. 1990). Figure~22 shows the effect
of changing the value of the dust emissivity. The smaller value of the
emissivity index implies that galaxies emit more flux at longer
wavelengths. As a consequence, the predicted CIB has more power at
$\lambda\ge$250~$\mu$m relative to the $\epsilon$=2 case, but fails to
reproduce the intensity at $\lambda<$200~$\mu$m. As discussed earlier,
the FIR energy distribution of galaxies is more complex than the
single temperature blackbody adopted here. Thus, the effect on the CIB
of changing the dust emissivity index should be regarded only as
indicative.

\subsubsection{The Metal Yield}

The metal yields adopted throughout the paper are close to the median
value observed in our Galaxy (Pagel 1987), although higher values
cannot be excluded. The impact of higher $\alpha_Z$ values on our
results are straightforward: higher yields mean larger mass in metals.
We have seen in Section~3 that changing the true yield $\alpha_Z$ from
0.7 to 1.0 generally increases the UV opacities.  If the final
metallicity is constrained to Z$_{\odot}$, outflows must be present,
implying that a larger initial galaxy mass is needed to outflow the
larger mass in metals. Larger initial galaxy masses imply larger
opacities at all redshifts. In numbers, by changing $\alpha_Z$ from
0.7 to 1.0, the absolute variation of 1500~\AA~ emergent-to-total UV
light is $\delta$=($-$0.08,$-$0.10) and the absolute variation of 2800~\AA~
emergent-to-total UV light is $\delta$=$-$0.07, implying
that the UV opacities are on average 40--50\% higher.

\section{Summary and Conclusions}

We have built a model for the cosmic evolution of the dust opacity in
galaxies in order to reconstruct the {\it intrinsic} SFR history of the
Universe, free from the effects of dust obscuration. The model uses
two basic constraints: the observed SFR density as a function of
redshift, derived from UV observations of galaxies (Lilly et al. 1996,
Madau et al. 1996, Connolly et al. 1997), and the Cosmic Infrared
Background observed by COBE (Hauser et al. 1998, Fixsen et al. 1998).
Additional constraints are provided by H$\alpha$, FIR and radio surveys 
of the local Universe and by the properties of the Lyman-break galaxies 
(e.g., Steidel et al. 1996, Giavalisco et al. 1996) and of the DLAs 
(e.g., Pettini et al. 1997a, 1998b) at high redshift. 

The model converges to multiple solutions for the intrinsic SFR
density (Figure~3) and the available observational constraints are
still uncertain enough that no secure choice can be made for any of
the solutions. The solutions can be separated into two main
classes. The first class is represented by our Model~2, for which the
intrinsic SFR has basically the same shape as the observed, UV-derived
SFR, with a peak at z$\simeq$1.2; this solution corresponds to modest
dust attenuations (2$\times$) of the UV emission from galaxies at high
redshift, but larger attenuations ($\sim$4$\times$) at
z$\le$1. According to this model, about 3/4 of the UV emission from
galaxies is hidden by dust in the low-z Universe, while this fraction
decreases to about 1/2 at z$>$1.2. The intrinsic SFR density of
Model~2 has the same intensity at z$\sim$3 and z$\sim$0.3. The second
class of solutions is represented by Model~3, for which the intrinsic
SFR density is constant at high redshift and declines at z$<$1.2 with
the same trend of the observed SFR density. Model~3 corresponds to
relatively large dust attenuations of the UV emission from galaxies at
high redshift ($\sim$5--6$\times$ at z$\sim$3) and smaller
attenuations ($\sim$3$\times$) at z$\le$1. In other words, according
to Model~3 local galaxies have on average enough dust opacity to
`hide' about 2/3 of the UV emission, while only $\sim$1/5 of the star
formation emerges at UV wavelengths from z$\sim$3 galaxies. The two
classes are mainly differentiated by the behavior of the intrinsic SFR
at z$>$1, in one case declining as the redshift increases (Model~2),
and in the other case constant with redshift (Model~3). Because of its
shape, the solution Model~3 resembles the `monolithic collapse' case
for galaxy formation, and currently this solution satisfies the
available observational constraints better than Model~2.

The various SFR$_{int}$ we derive imply rather different distributions
of the flux contribution to the CIB as a function of redshift. Model~3
predicts that most of the CIB flux at 850~$\mu$m originates at
redshift greater than z$\sim$2, while Model~2 predicts equal flux
contributions from all galaxies at z$>$1. SCUBA observations should
therefore be able to discriminate between the two models, once better
redshift placements for the detected sources are secured.

It is worth noting that the input assumptions we have used to derive
the self-consistent solutions tend to minimize the amount of dust
produced (small metal yield, evolving dust/metal ratio) and its
obscuring effects (mixed dust/star geometry). We have seen in
section~4.2.5 that a 25\% larger value for the metal yield immediately
boosts up all UV opacities by $\sim$50\%. Thus, under certain
conditions galaxies may be even more opaque than what we have
modelled.

Modelling the evolution of galaxies and their properties suffers from
a number of uncertainties and unknowns about the physics of galaxy and
star formation. Many of these have been detailed in Sections~2 and
4.2. The stellar IMF brings a factor $\sim$2 uncertainty in the
derivation of the global SFR, because the low mass slope and cut-off
mass are not measured with sufficient accuracy. Most likely, the
uncertainty is in the direction of overestimating SFRs from the
measurements of massive stars emission. Another question mark is
whether the IMF has been constant with time or has changed as the
physical conditions in galaxies (energy, metal, dust contents)
changed. Linked to the IMF is the problem of the effective yield
$\alpha_Z$ (briefly discussed in section~4.2.5), which drives the
metal and dust production and which even in our Galaxy is still
uncertain by at least a factor 2 (Pagel 1987, 1998). Finally,
uncertainties in the baryonic mass fraction in galaxies introduce
another $\sim$50\% uncertainity in models.

In conclusion, simple models of galaxy evolution indicate that star
formation induces non negligible dust opacities at all redshifts
(A$_{1500}\sim$0.8--2 at z$>$1 and A$_{2800}\sim$1.2--1.5 at
z$\le$1). These opacities should be taken into account when
reconstructing the global star formation history of the Universe from
dust-sensitive observables. We emphasize that the attenuations derived
here are perfectly adequate to reproduce the intensity and the
spectral energy distribution of the CIB, arguing against the necessity
of larger dust opacities in galaxies.

\acknowledgments

The idea for this paper was born during the Aspen Winter Workshop on
``Universal Star Formation'' (January 1998), from discussions with
Rosemary Wyse, Mauro Giavalisco and Alvio Renzini.  D.C. acknowledges
the hospitality of the Observatories of the Carnegie Institution of
Washington, where part of this work was developed. The authors also
thank Mark Dickinson for elucidating the properties of the high
redshift galaxies.

{}

\clearpage
\begin{figure}
\plotone{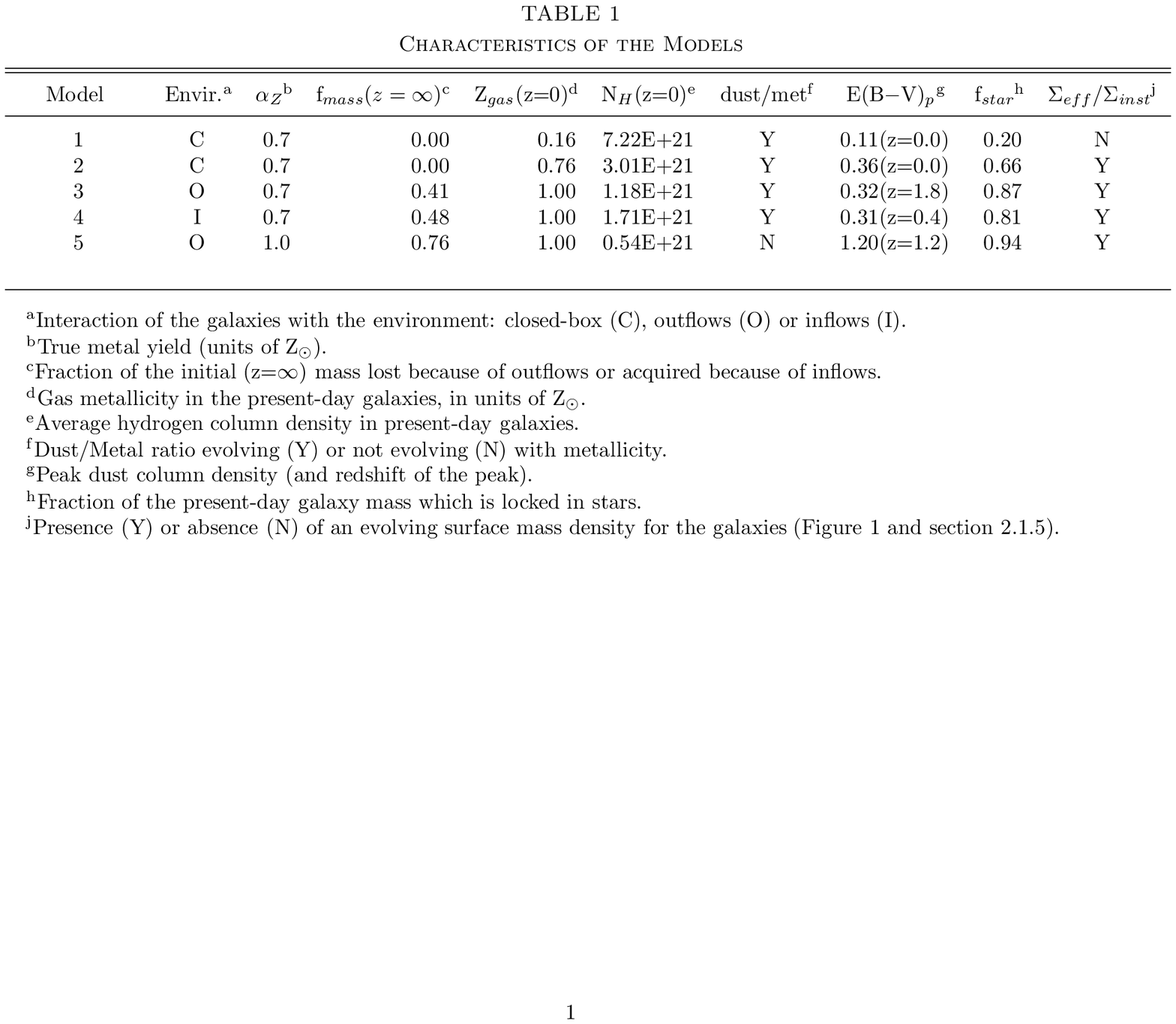}
\end{figure}

\clearpage
\begin{figure}
\plotone{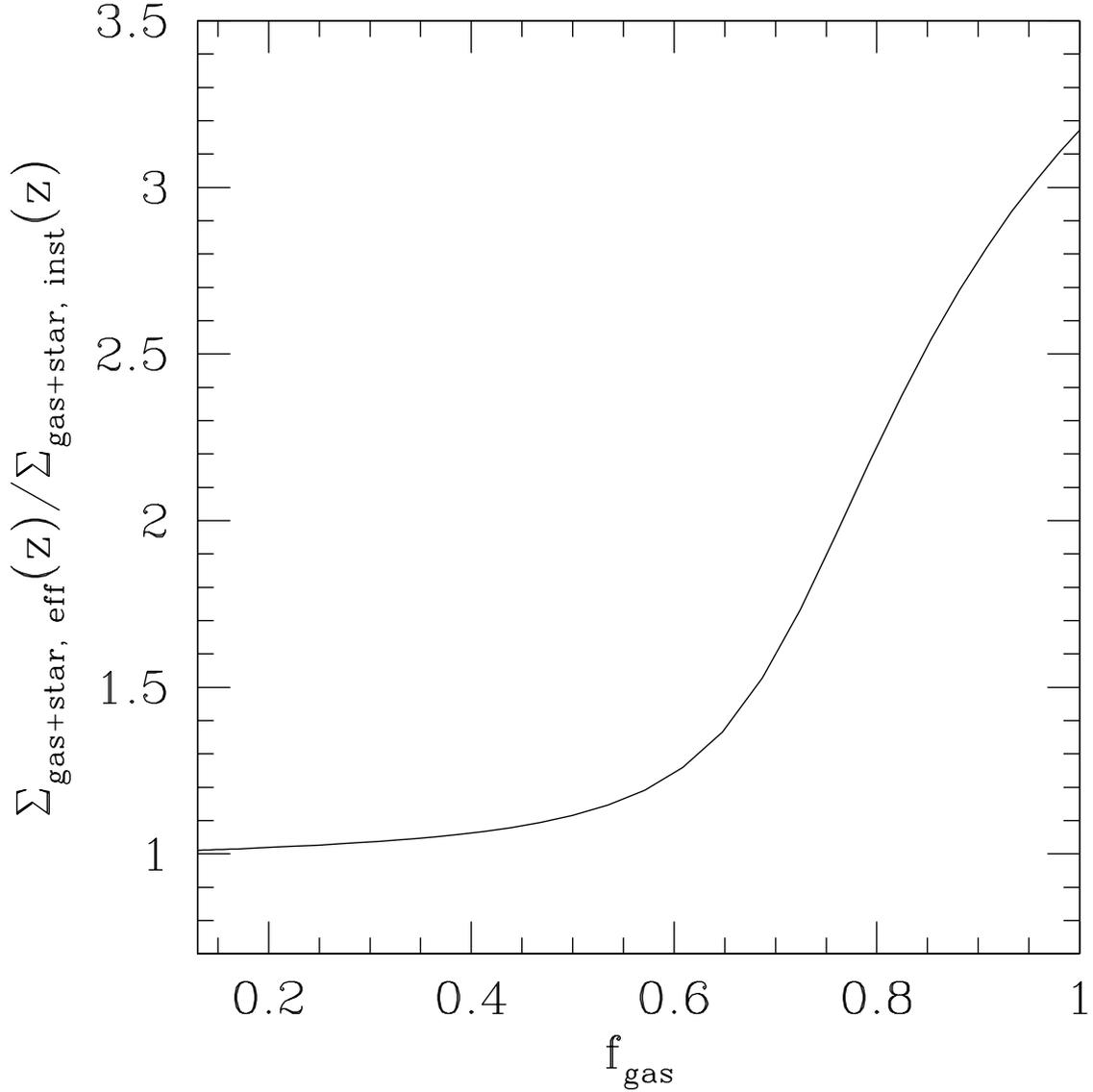}
\figcaption[ch_figure01.ps]{The ratio of the effective-to-instantaneous
surface mass density $\Sigma_{gas+star, eff}$(z)/$\Sigma_{gas+star,
inst}$(z), as a function of the gas fraction in the
galaxy. $\Sigma_{gas+star, inst}$(z) is the average surface mass
density at redshift z from the outflow/closed-box/inflow
models. $\Sigma_{gas+star, eff}$(z) is the effective surface mass
density observed in high redshift galaxies, possibly due to
concentrated star formation. The ratio $\Sigma_{gas+star,
eff}$(z)/$\Sigma_{gas+star, inst}$(z) is used to model the effective
dust column density and temperature in high redshift galaxies.}
\end{figure}

\begin{figure}
\plotone{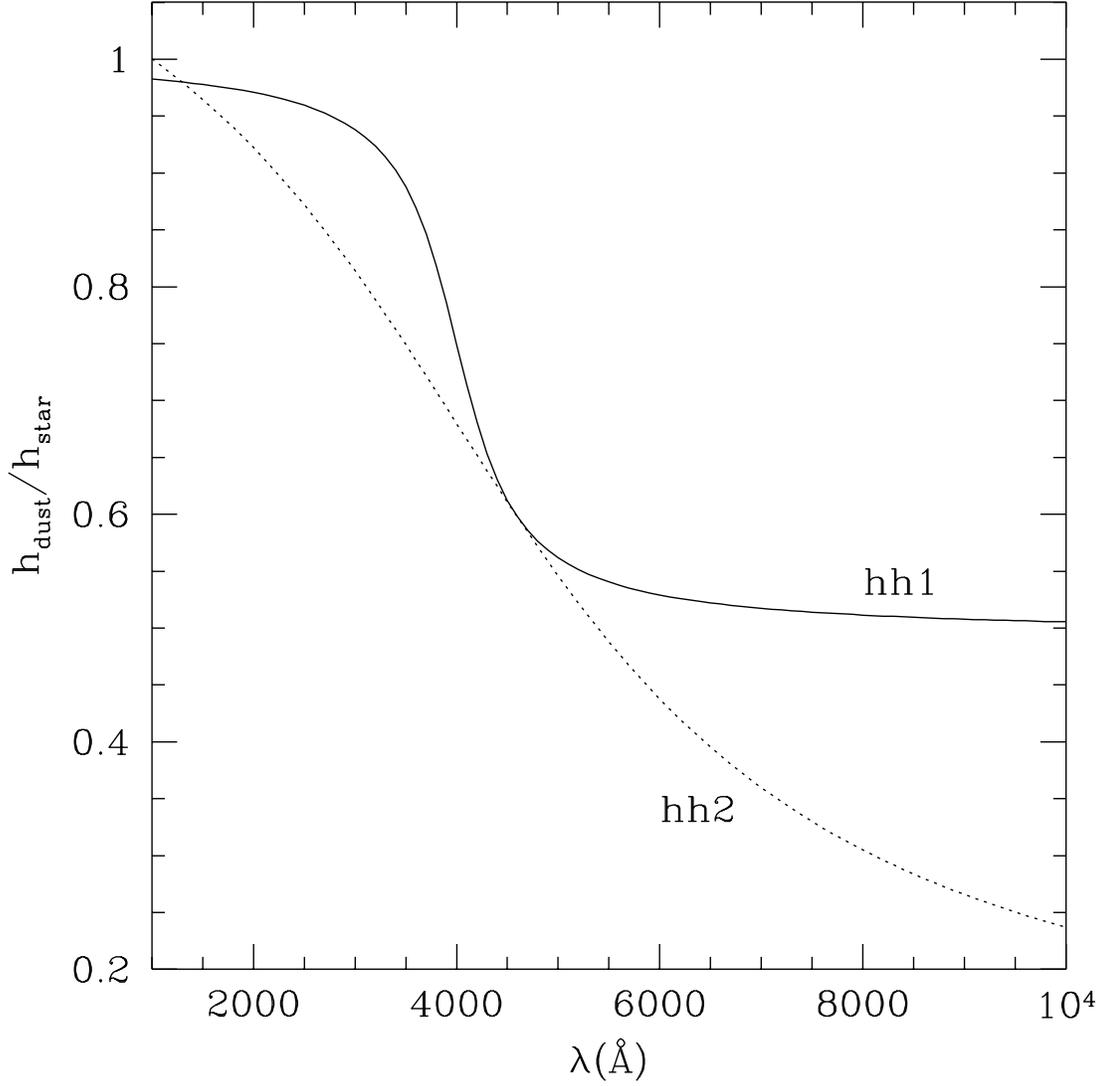}
\figcaption[ch_figure02.ps]{The two representations of the scaleheight 
ratio of dust to stars,
h$_{dust}$/h$_{star}$, adopted throughout the paper are shown here as
a function of wavelength. In the first representation, labelled `hh1',
the scaleheight ratio is $\sim$1 in the UV and is $\sim$0.5 at
optical wavelengths, with a sharp change around 4000~\AA.  The second
representation, labelled `hh2', accomodates a smoother change in the
scaleheight ratio, from $\sim$1 at 1000~\AA~ to $\sim$0.25 at
10000~\AA.}
\end{figure}

\clearpage
\begin{figure}
\plotone{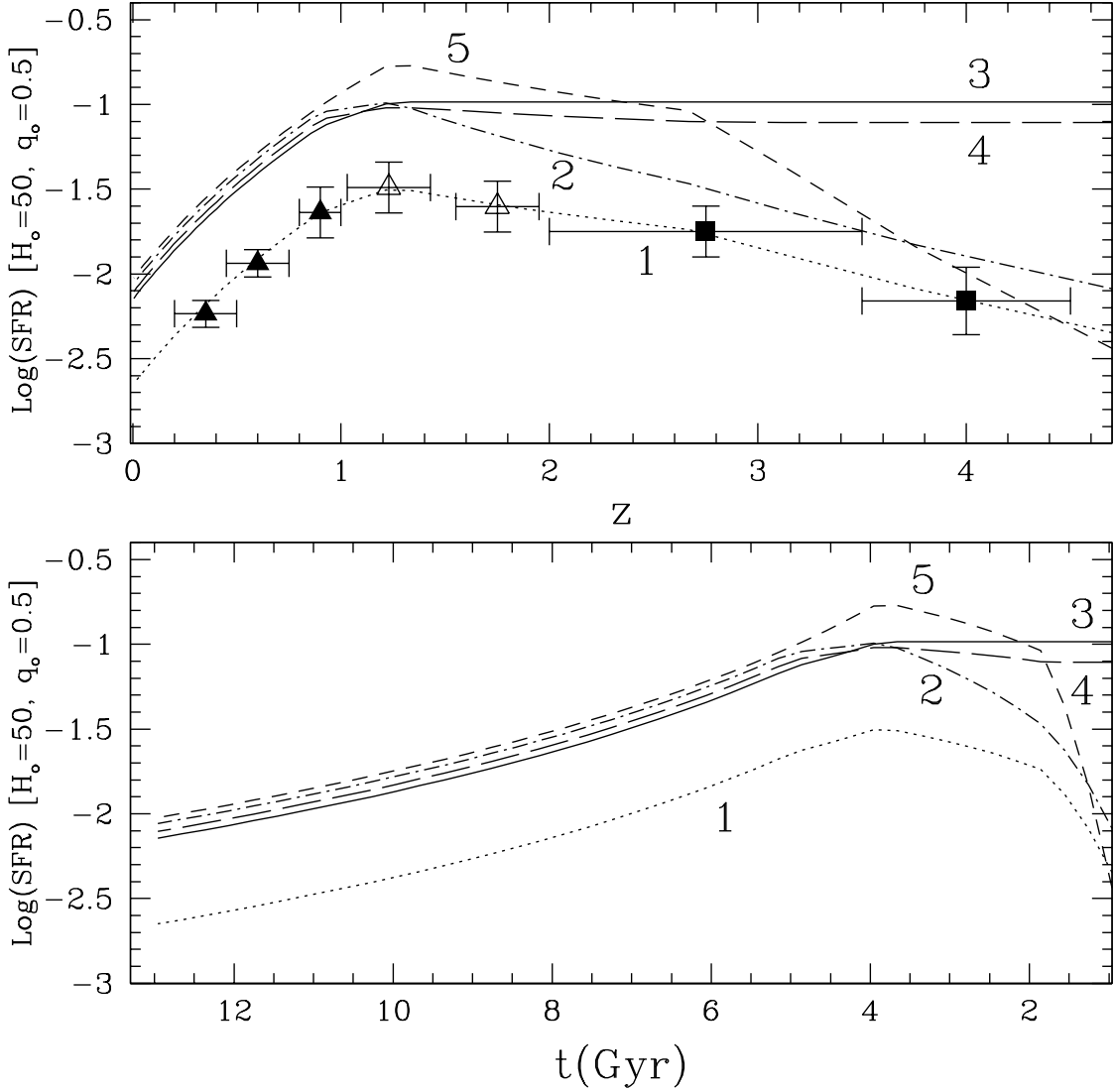}
\figcaption[ch_figure03.ps]{The observed and intrinsic star formation
rate densities SFR, in units of M$_{\odot}$~yr$^{-1}$~Mpc$^{-3}$, are 
shown as a function of both redshift (top panel) and time (bottom
panel). The observed SFR, SFR$_{obs}$ is labelled as Model~1 (dotted
line), and represents a smooth interpolation of the data points.  The
data are from the observed UV flux densities reported by Lilly et
al. (1996, filled triangles) and Connolly et al. (1997, empty
triangles) at 2800~\AA, and by Madau et al. (1998, filled squares) at
1500~\AA. The other four models are the intrinsic star formation
rate densities, SFR$_{int}$, which are solutions of our iterative 
procedure; the solutions
are labelled Model~2 through 5 and their main characteristics are
listed in Table~1. Models~2 (dot-dashed line), 3 (continuous line),
and 4 (long-dashed line) are solutions for closed-box, outflows and
inflows galaxy models, respectively. Model~5 is obtained from extreme
choices of some of the model's parameters (see text). In all cases the
low redshift SFR$_{int}$ is similar in amplitude, while most of the
difference occurs at z$>$1, namely at ages of the Universe $<$4~Gyr
(H$_o$=50~km/s/Mpc and q$_o$=0.5.).}
\end{figure}

\begin{figure}
\plotone{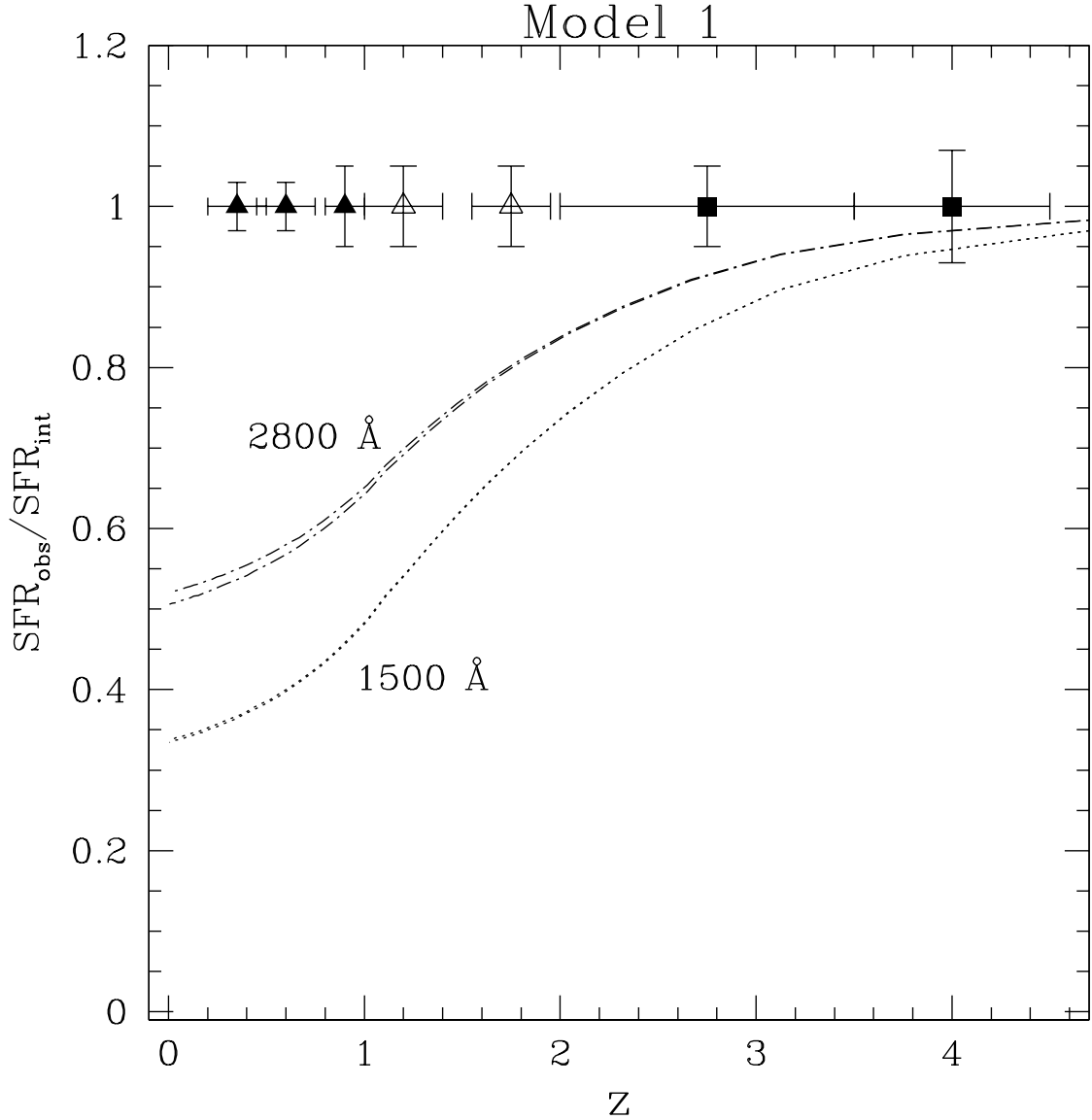}
\figcaption[ch_figure04.ps]{The ratio between the observed and
intrinsic star formation rate density, SFR, is shown as a function of
redshift. The data points (filled and empty triangles and filled
squares) show the ratio SFR$_{obs}$/SFR$_{int}$ using the SFR$_{obs}$
of Figure~3. Here the ratio is unity because SFR$_{int}$=SFR$_{obs}$
by construction in Model~1. The continuous lines are the
emerging-to-total radiation predicted by the input star formation
rate, SFR$_{int}$, at 2800~\AA~ (dash-dotted line) and
1500~\AA~(dotted line), respectively.  At each fixed wavelength, two
lines are shown and correspond to the two adopted scaleheight ratios,
h$_{dust}$/h$_{star}$ (Figure~2): the lower values of the
emerging-to-total radiation correspond to hh1, while the larger values
correspond to hh2. For SFR$_{int}$ to be a solution of the model, the
SFR$_{obs}$/SFR$_{int}$ points must lay on top of the opacity curve at
the appropriate wavelength. It is clear that Model~1 is an
unacceptable solution.}
\end{figure}

\begin{figure}
\plotone{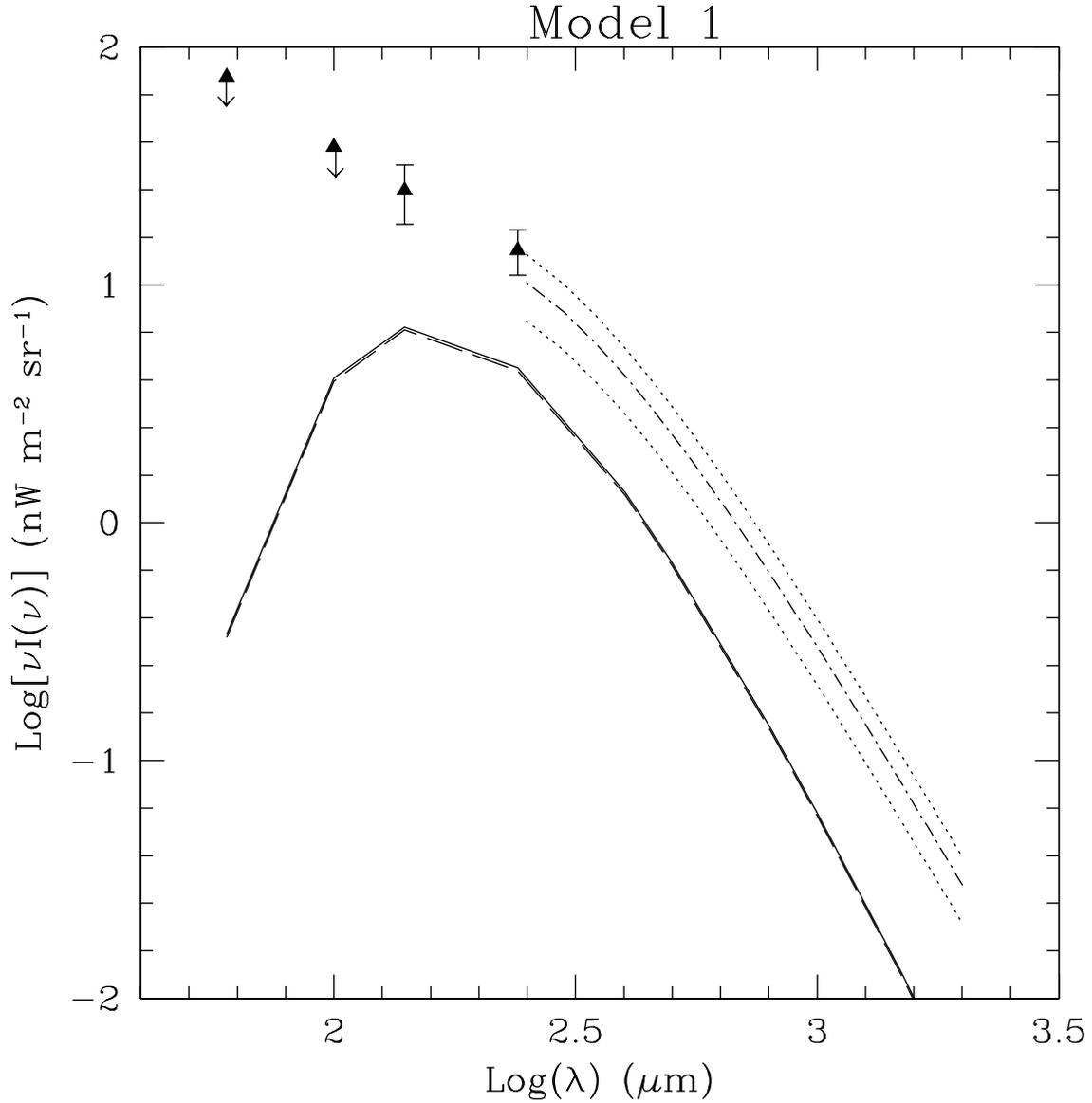}
\figcaption[ch_figure05.ps]{The FIR background predicted by Model~1 is
compared with the data from COBE DIRBE and FIRAS. The continuous line
is the model prediction for the dust/star scaleheight ratio hh1 and
the long-dashed line is the model prediction for hh2 (see
Figure~2). The DIRBE data are indicated as filled triangles with
1~$\sigma$ error bars; the values at 60~$\mu$m and 100~$\mu$m are
upper limits. The FIRAS values are reported as the best fit to the
data given in Fixsen et al. (1998, dot-dashed line) together with the
fiducial 1$\sigma$ error bar (dotted lines). Model~1 produces only
$\sim$25\%~ of the observed FIR flux, with no significant variation
between the two dust/star scaleheight ratios.}
\end{figure}

\begin{figure}
\plotone{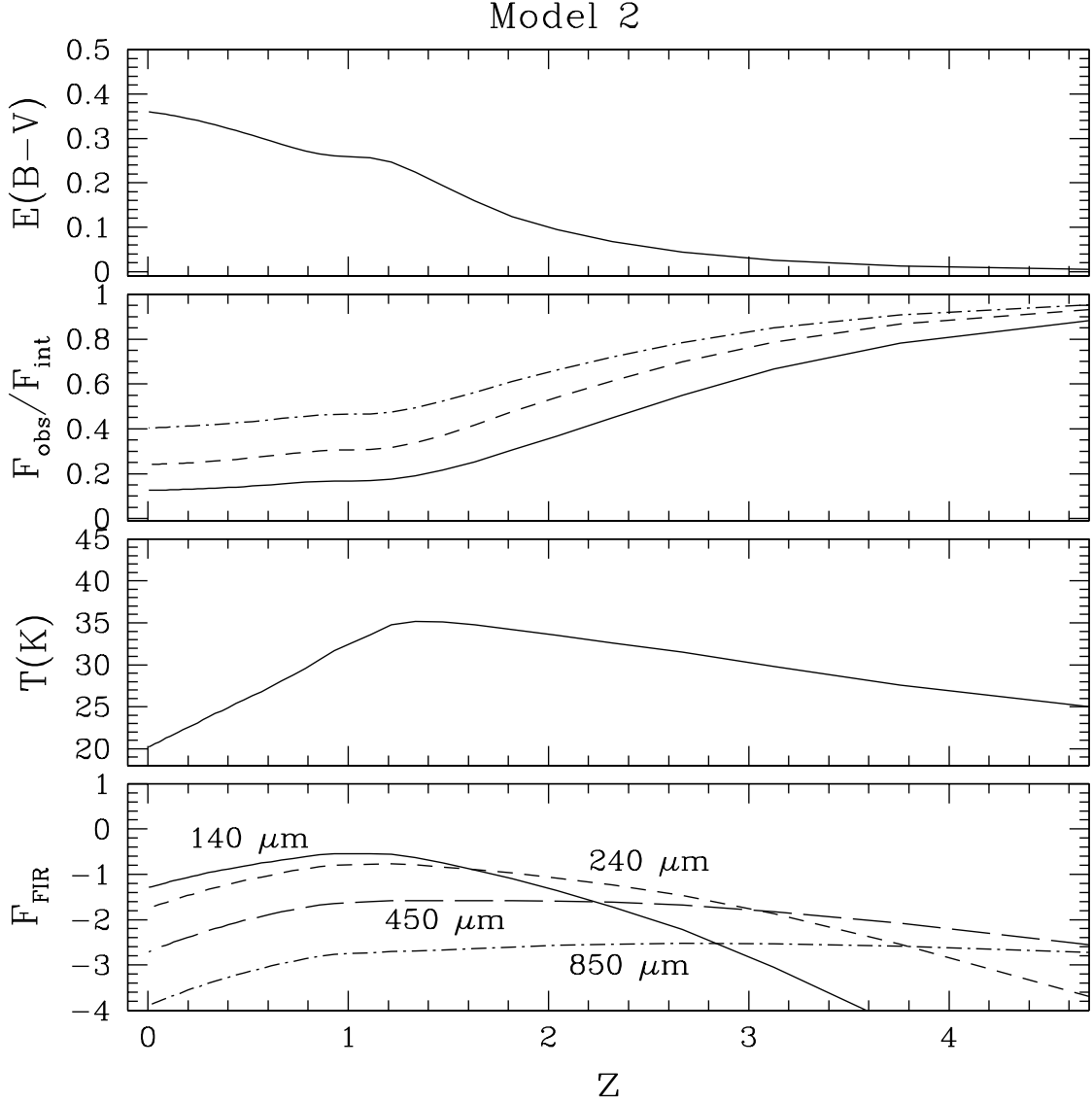}
\figcaption[ch_figure06.ps]{The evolution of the dust column density, the
opacity and the FIR emission as a function of the redshift z for the
SFR$_{int}$ of Model~2 (cf. Figure~3).  The first panel from the top
shows the evolution of the dust column density E(B$-$V). The second
panel shows the ratio of emerging-to-total radiation for a few
representative UV-B wavelengths in the galaxy's restframe: 1500~\AA~
(solid line), 2800~\AA~ (dashed line), and 4400~\AA~ (dot-short dash
line). The third panel shows the evolution of the dust temperature in
K. The fourth panel shows the contributing flux to the CIB, in
arbitrary units, at selected wavelengths in the observer's restframe,
140~$\mu$m, 240$\mu$m, 450~$\mu$m and 850~$\mu$m. The dust/star
scaleheight ratio adopted in this figure is hh1 (Figure~2).}
\end{figure}

\begin{figure}
\plotone{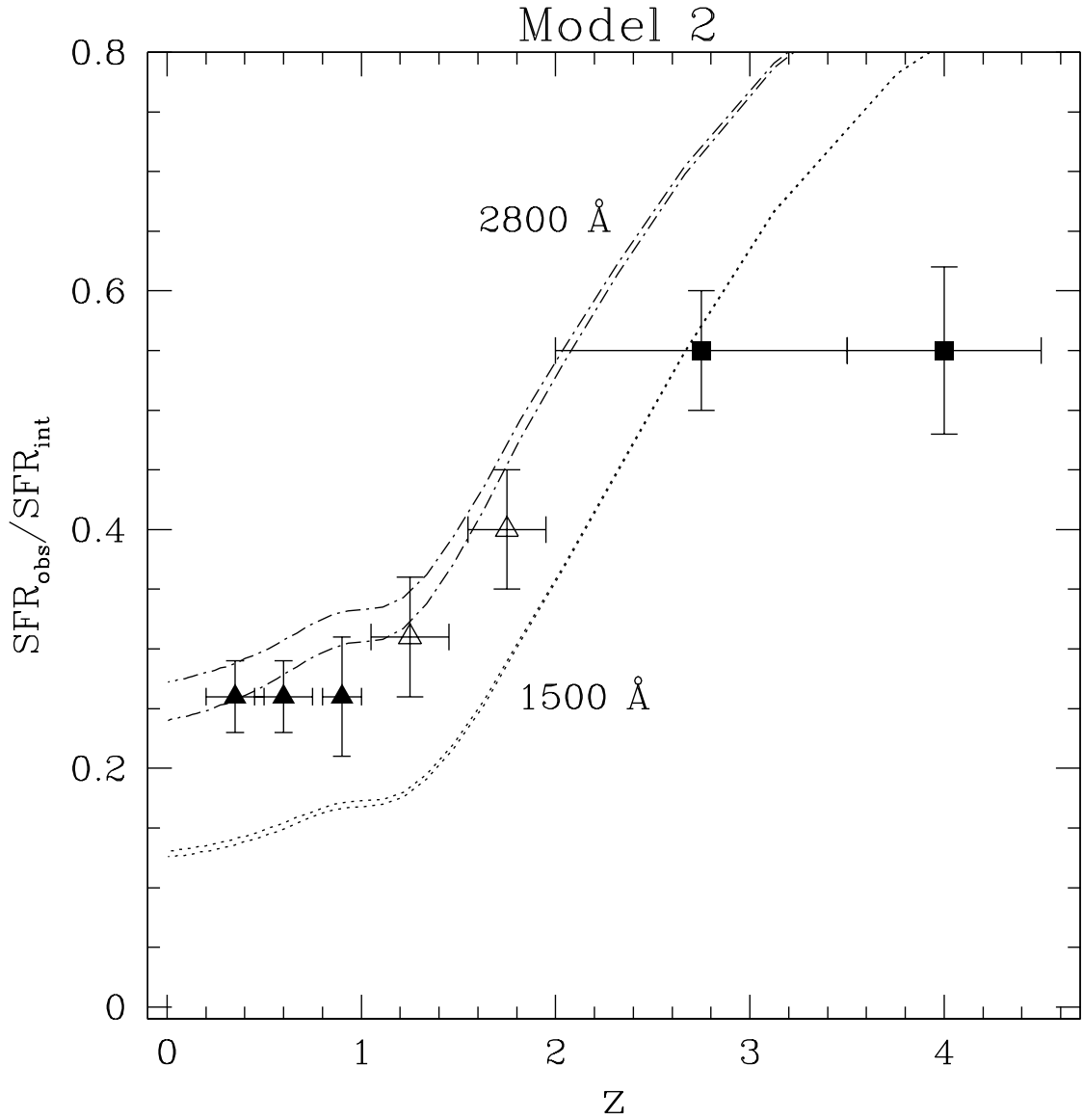}
\figcaption[ch_figure07.ps]{As Figure~4, for the SFR(z) of
Model~2. The ratio SFR$_{obs}$/SFR$_{int}$ at z$<$2, which is
equivalent to the emergent-to-total light at 2800~\AA, overlaps with
the predicted opacity curve at the same wavelength. The same happens
for the SFR$_{obs}$/SFR$_{int}$ point at z=2.75, which is equivalent
to the emergent-to-total light at 1500~\AA. This figure shows that 
SFR$_{int}$(z) of Model~2 is a solution of our self-consistent procedure as 
the predicted UV opacities account for SFR$_{obs}$(z).}
\end{figure}

\begin{figure}
\plotone{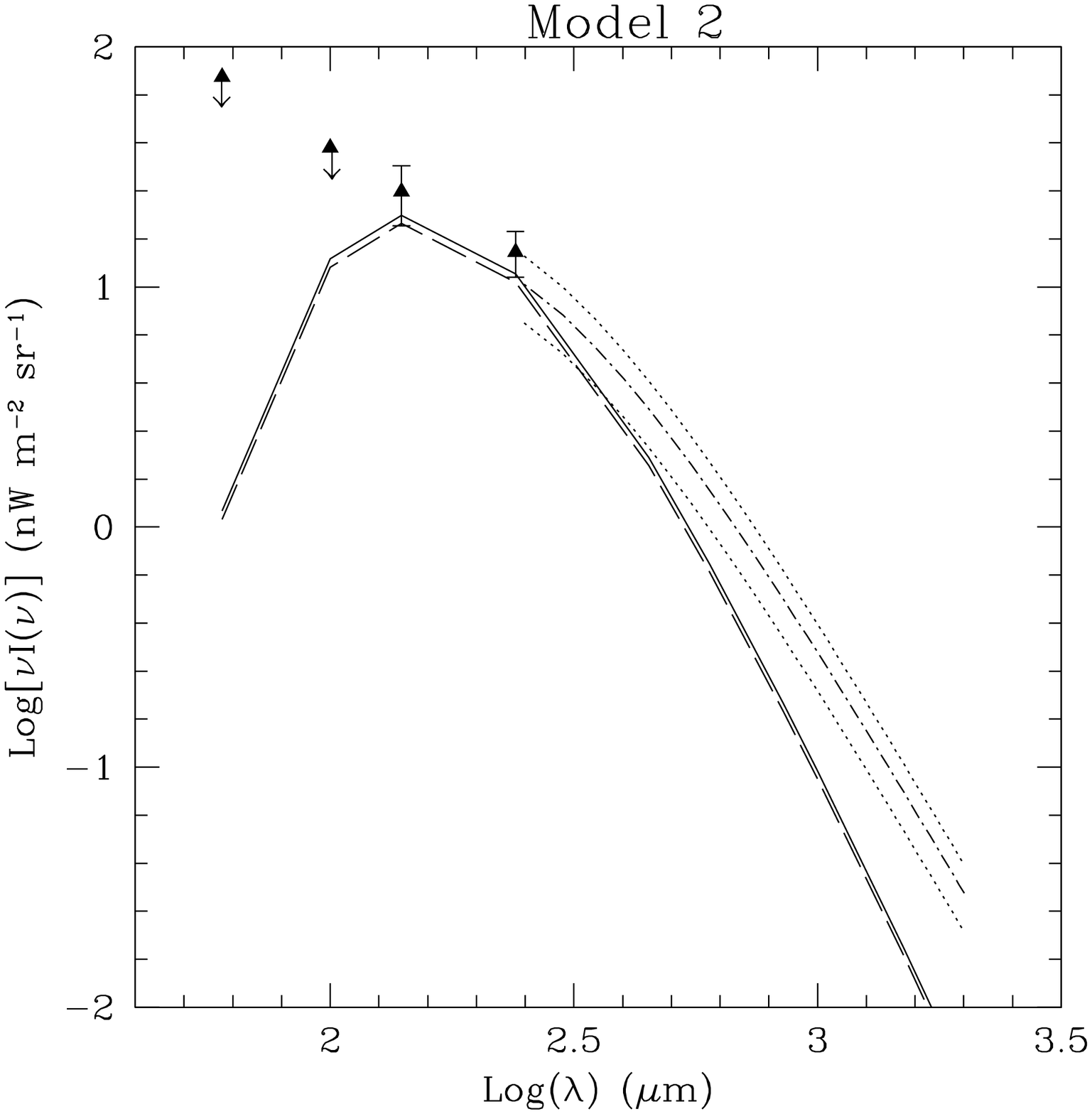}
\figcaption[ch_figure08.ps]{The FIR background predicted by the
SFR$_{int}$ of Model~2 is compared with the data from COBE DIRBE and
FIRAS. The symbols and lines are as in Figure~5.}
\end{figure}

\begin{figure}
\plotone{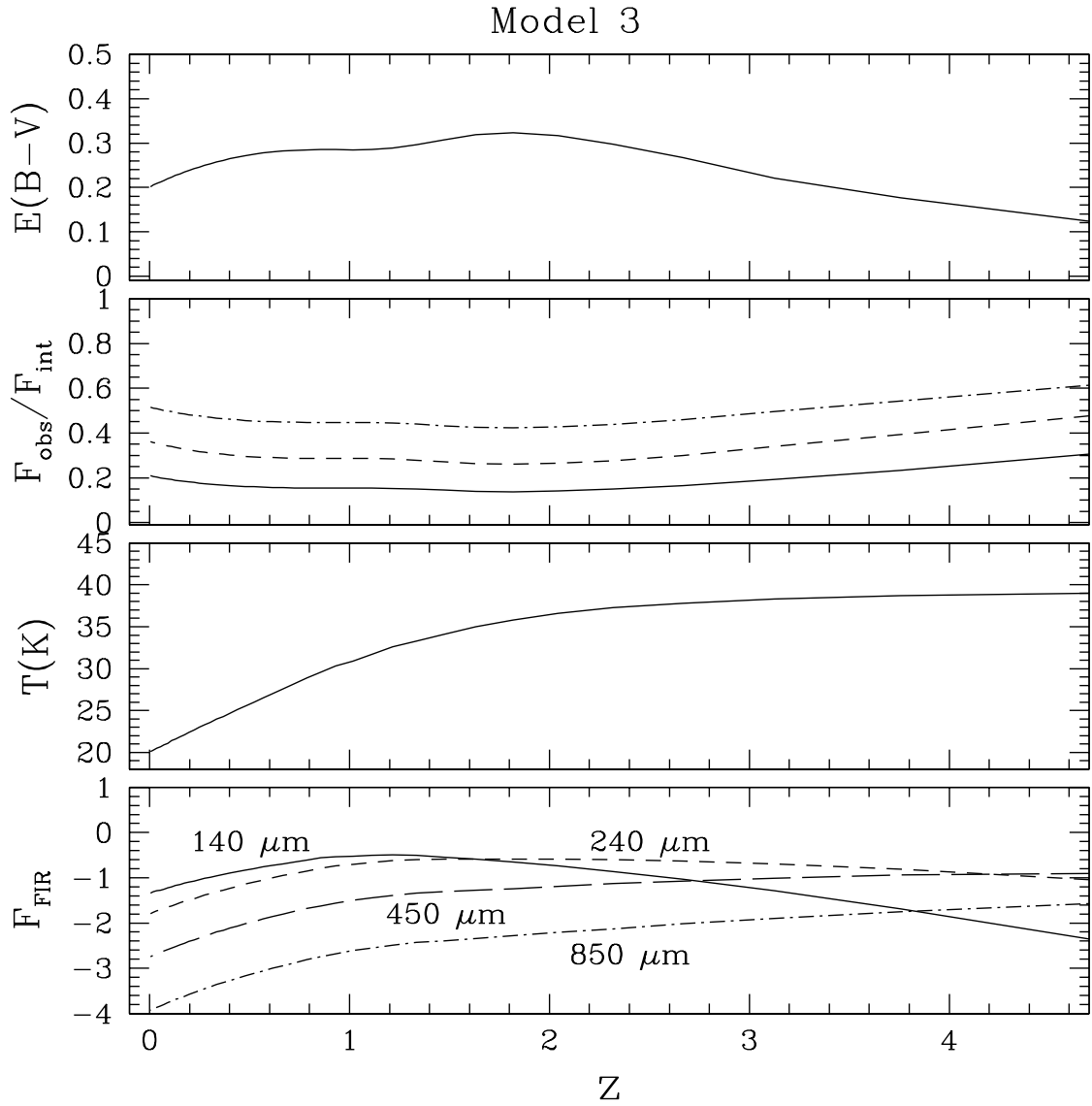}
\figcaption[ch_figure09.ps]{As Figure~6, for SFR$_{int}$ of Model~3.}
\end{figure}

\begin{figure}
\plotone{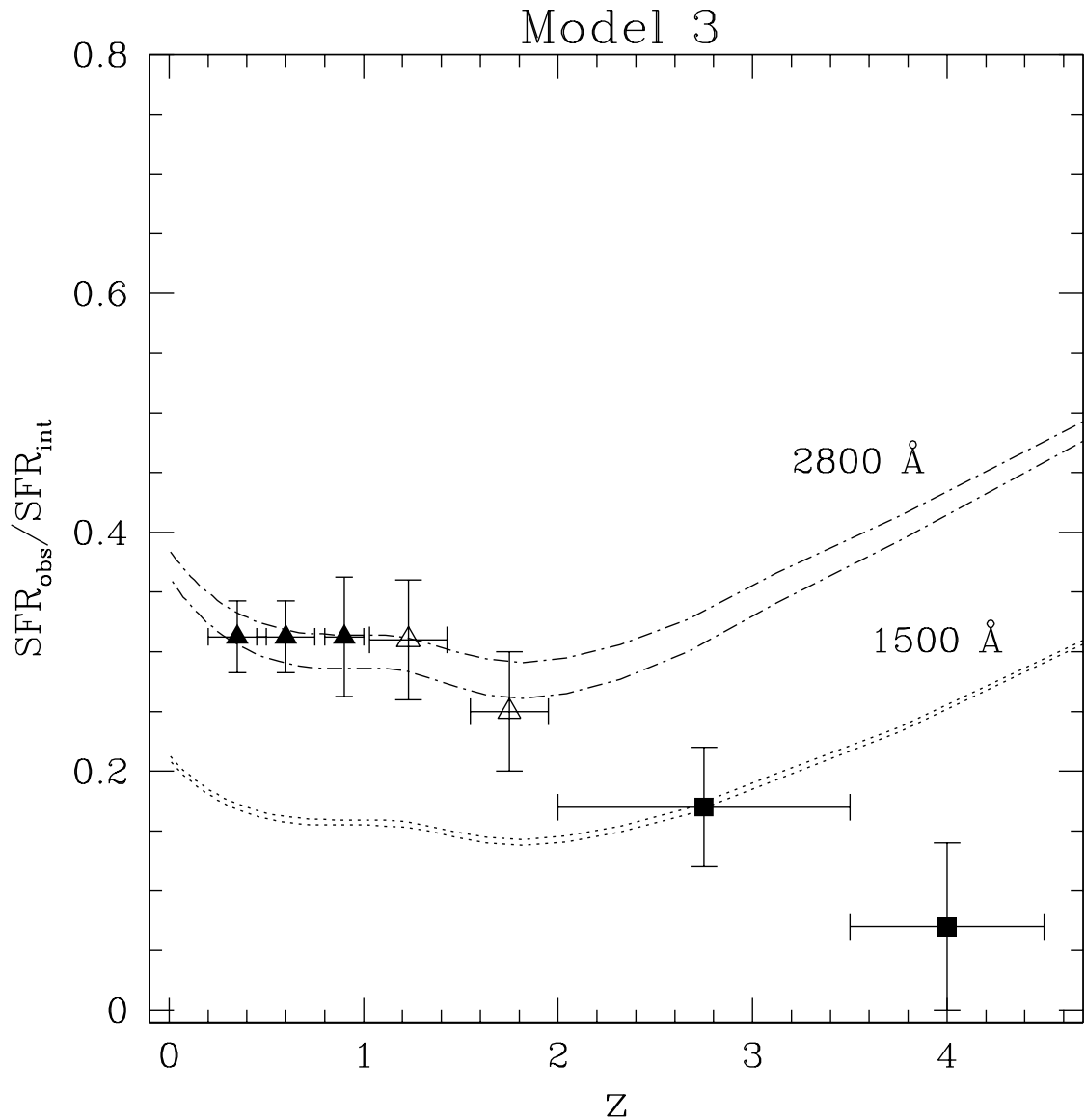}
\figcaption[ch_figure10.ps]{As Figure~4, for the SFR(z) of Model~3. The
agreement between SFR$_{int}$/SFR$_{obs}$ and predicted UV opacities
implies that SFR$_{int}$ of Model~3, as Model~2, is also a solution of
our self-consistent procedure.}
\end{figure}

\begin{figure}
\plotone{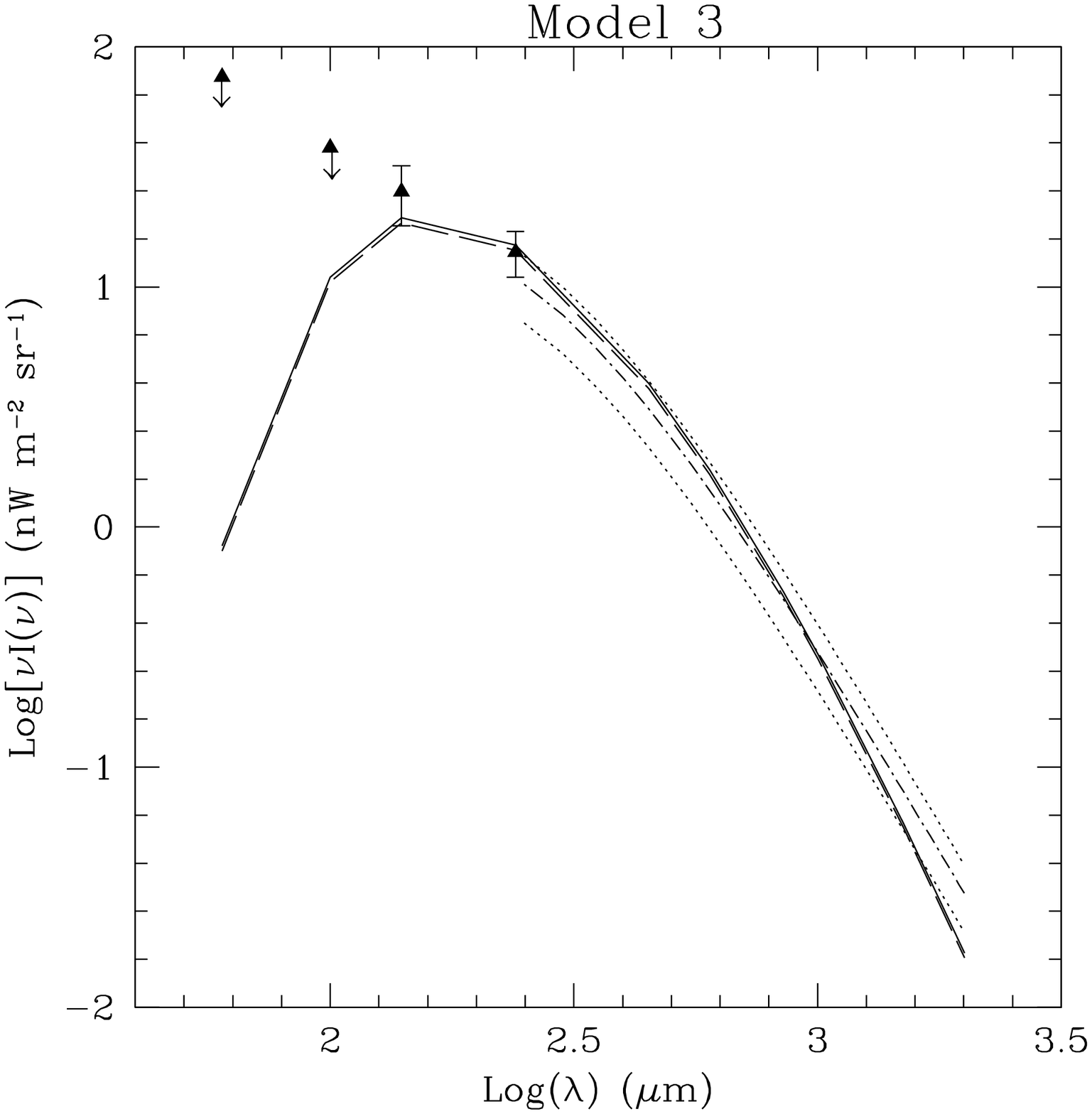}
\figcaption[ch_figure11.ps]{The FIR background predicted by the
SFR$_{int}$ of Model~3 is compared with the data from COBE DIRBE and
FIRAS. The symbols and lines are as in Figure~5.}
\end{figure}

\begin{figure}
\plotone{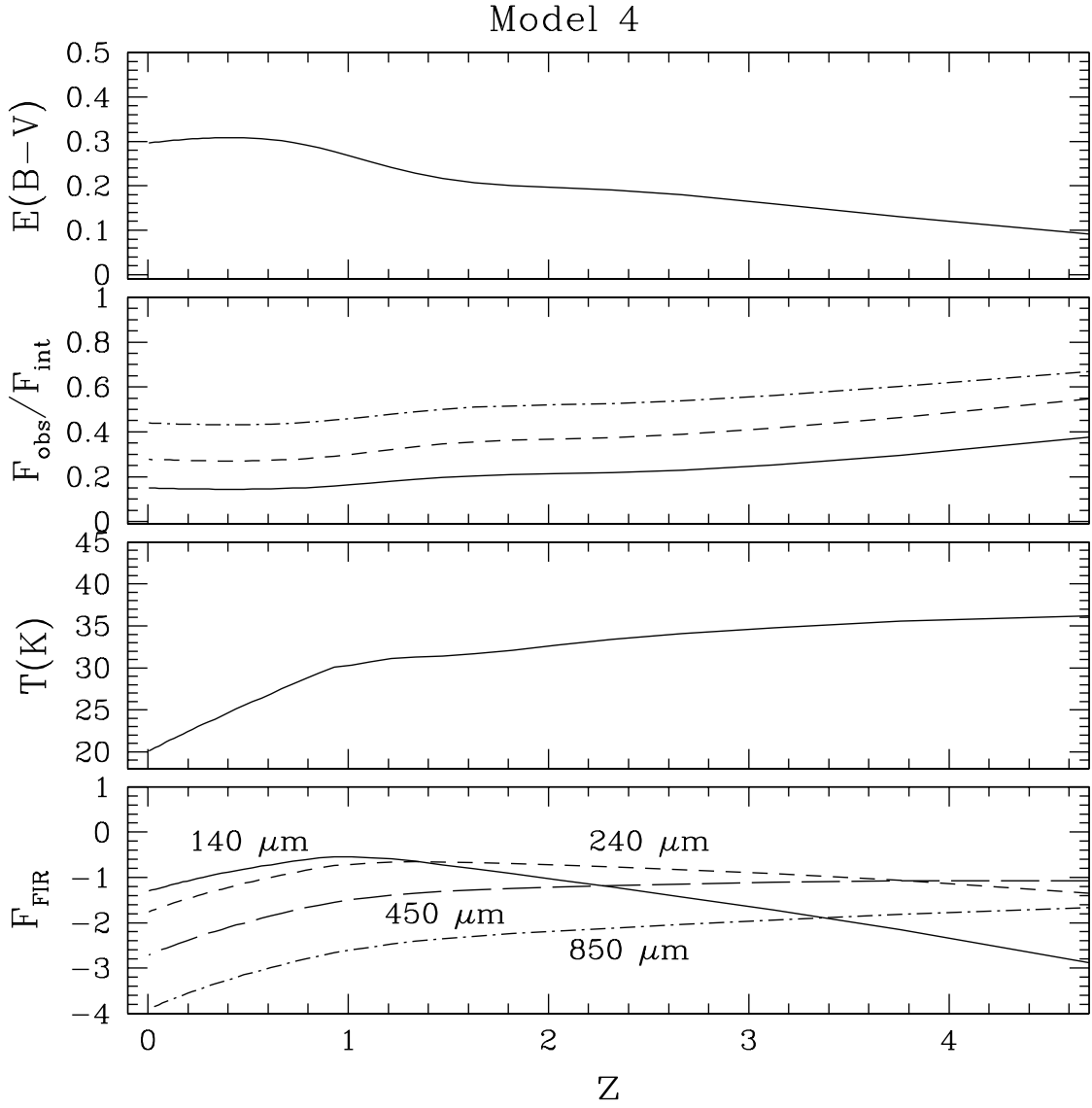}
\figcaption[ch_figure12.ps]{As Figure~6, now for SFR$_{int}$ of Model~4.}
\end{figure}

\begin{figure}
\plotone{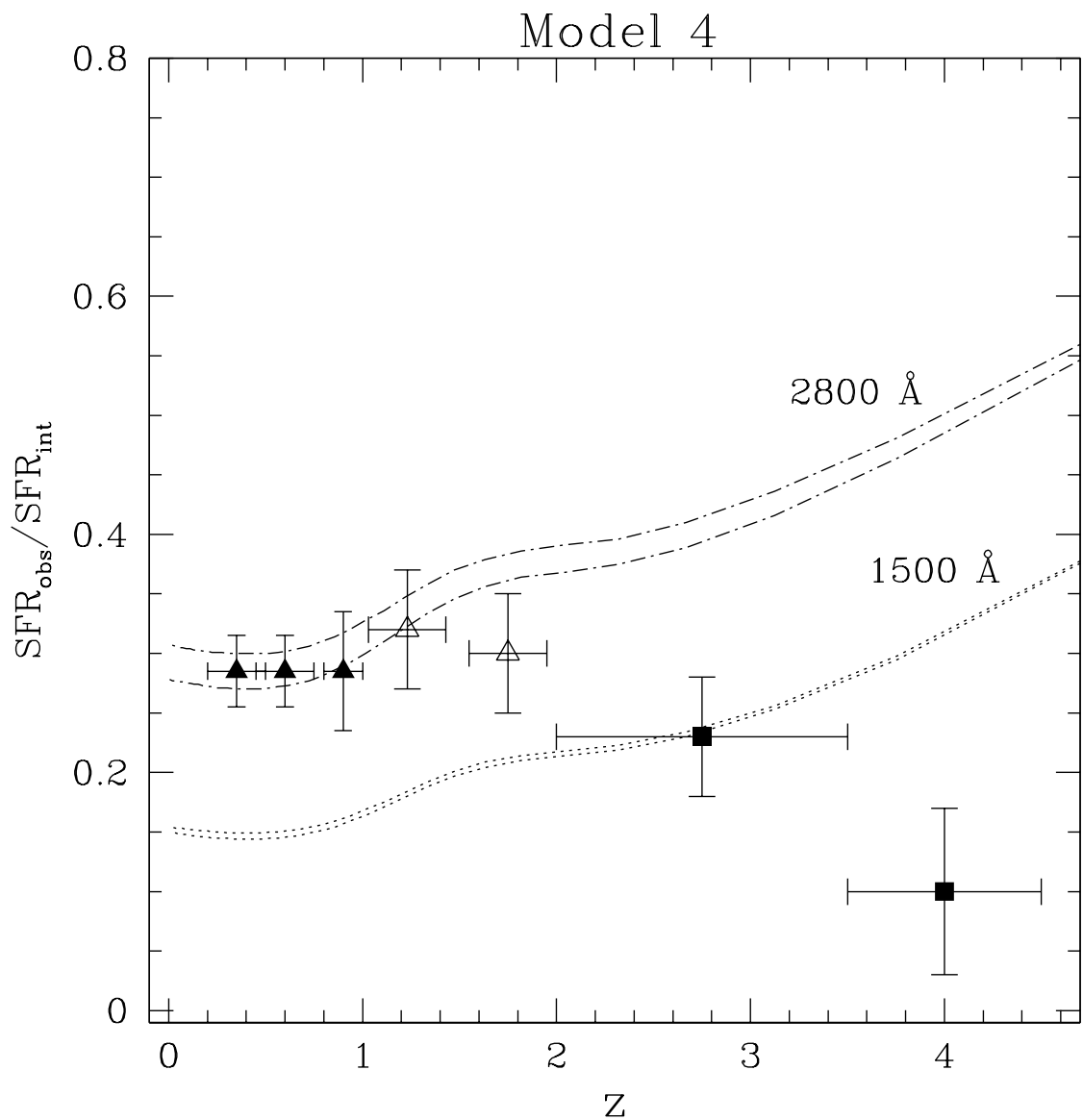}
\figcaption[ch_figure13.ps]{As Figure~4, for the SFR(z) of Model~4. The 
agreement between SFR$_{int}$/SFR$_{obs}$ and predicted UV opacities 
implies that SFR$_{int}$ of Model~4 is another acceptable solution of the 
self-consistent procedure.}
\end{figure}

\begin{figure}
\plotone{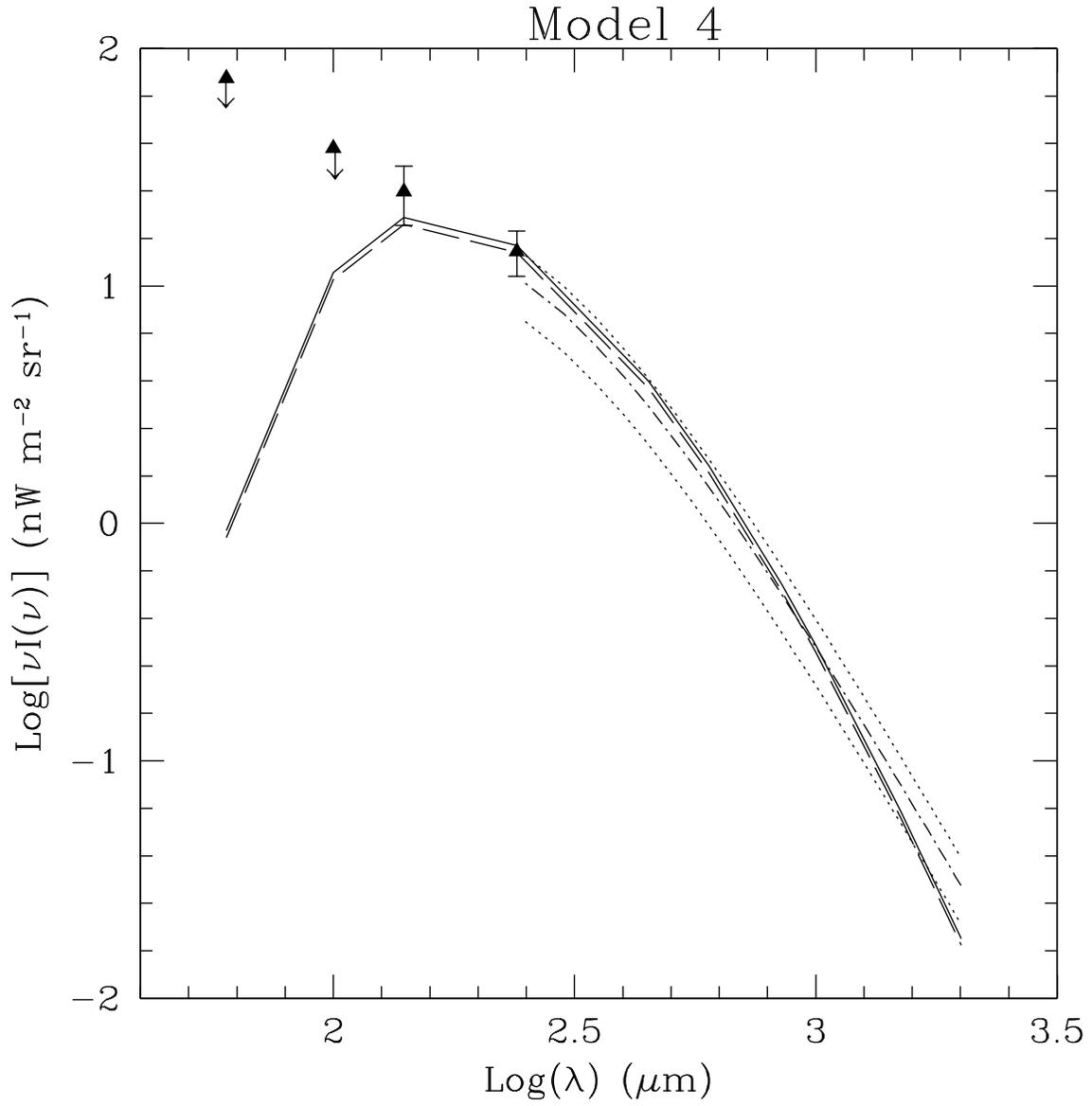}
\figcaption[ch_figure14.ps]{The FIR background predicted by Model~4 is
compared with the data from COBE DIRBE and FIRAS. The symbols and lines 
are as in Figure~5.}
\end{figure}

\begin{figure}
\plotone{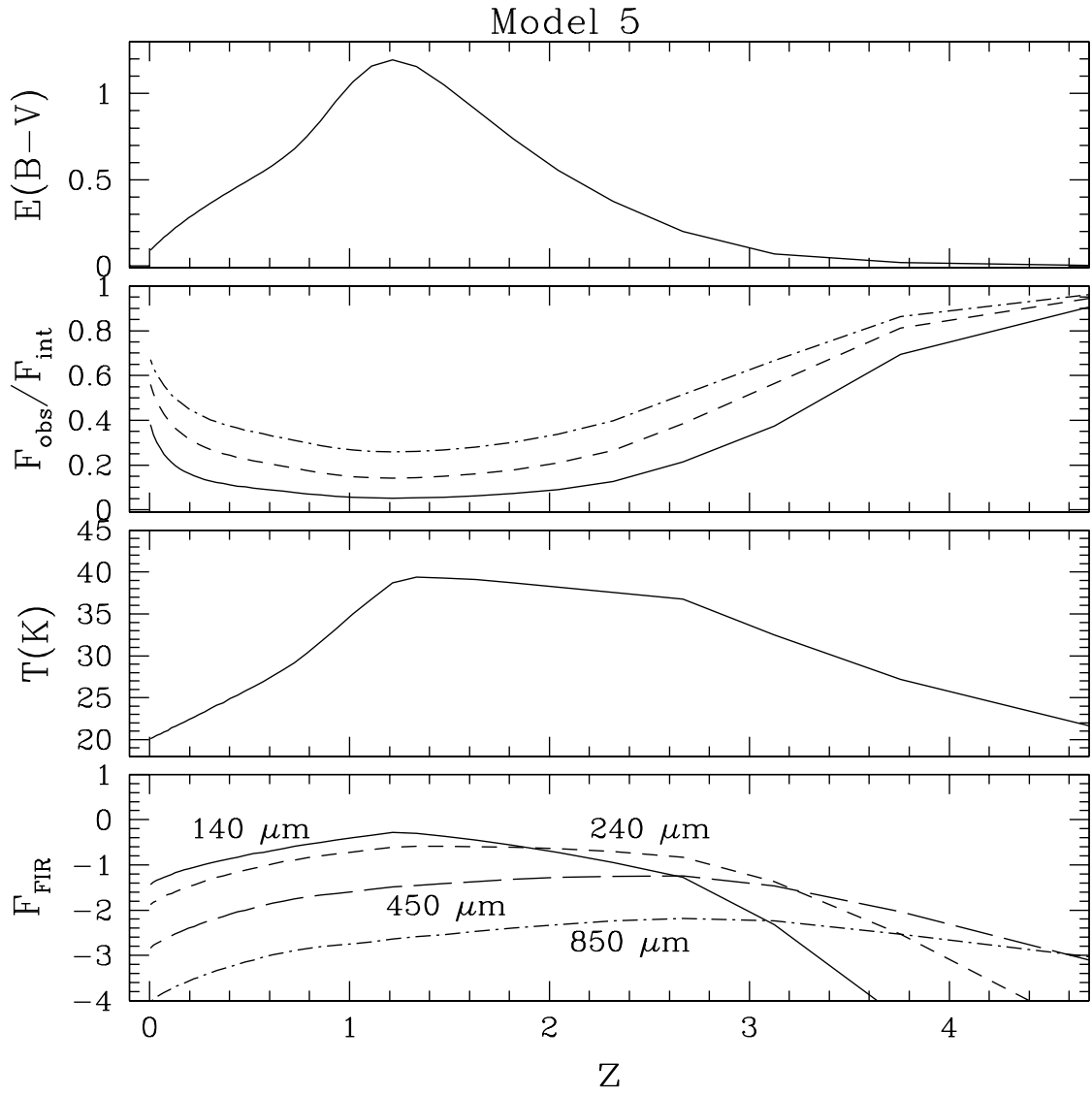}
\figcaption[ch_figure15.ps]{As Figure~6, for SFR$_{int}$ of Model~5.}
\end{figure}

\begin{figure}
\plotone{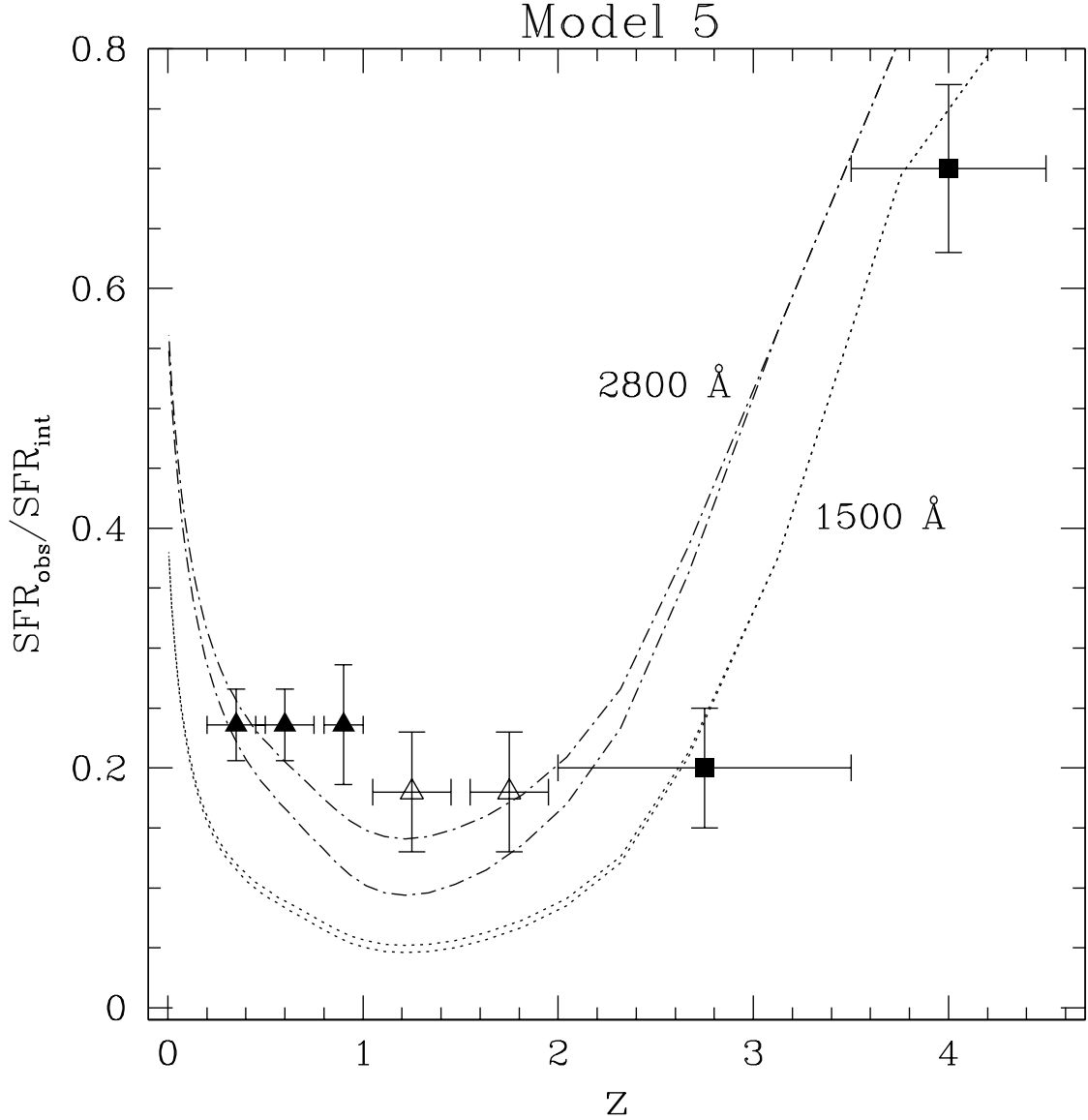}
\figcaption[ch_figure16.ps]{As Figure~4, for the SFR(z) of Model~5. The
SFR$_{obs}$/SFR$_{int}$ points at z=2.75 and at z=4, which are
equivalent to the emergent-to-total light at 1500~\AA, overlap with
the predicted opacity curve at the same wavelength. The
SFR$_{obs}$/SFR$_{int}$ at z$<$2, however, does not reproduce the
curve of the emergent-to-total light at 2800~\AA\ as well as the
previous three models do. This figure shows that the SFR$_{int}$ of
Model~5 is almost a solution for the iterative procedure, but the
SFR$_{int}$/SFR$_{obs}$ does not completely account for the predicted
UV opacities.}
\end{figure}

\begin{figure}
\plotone{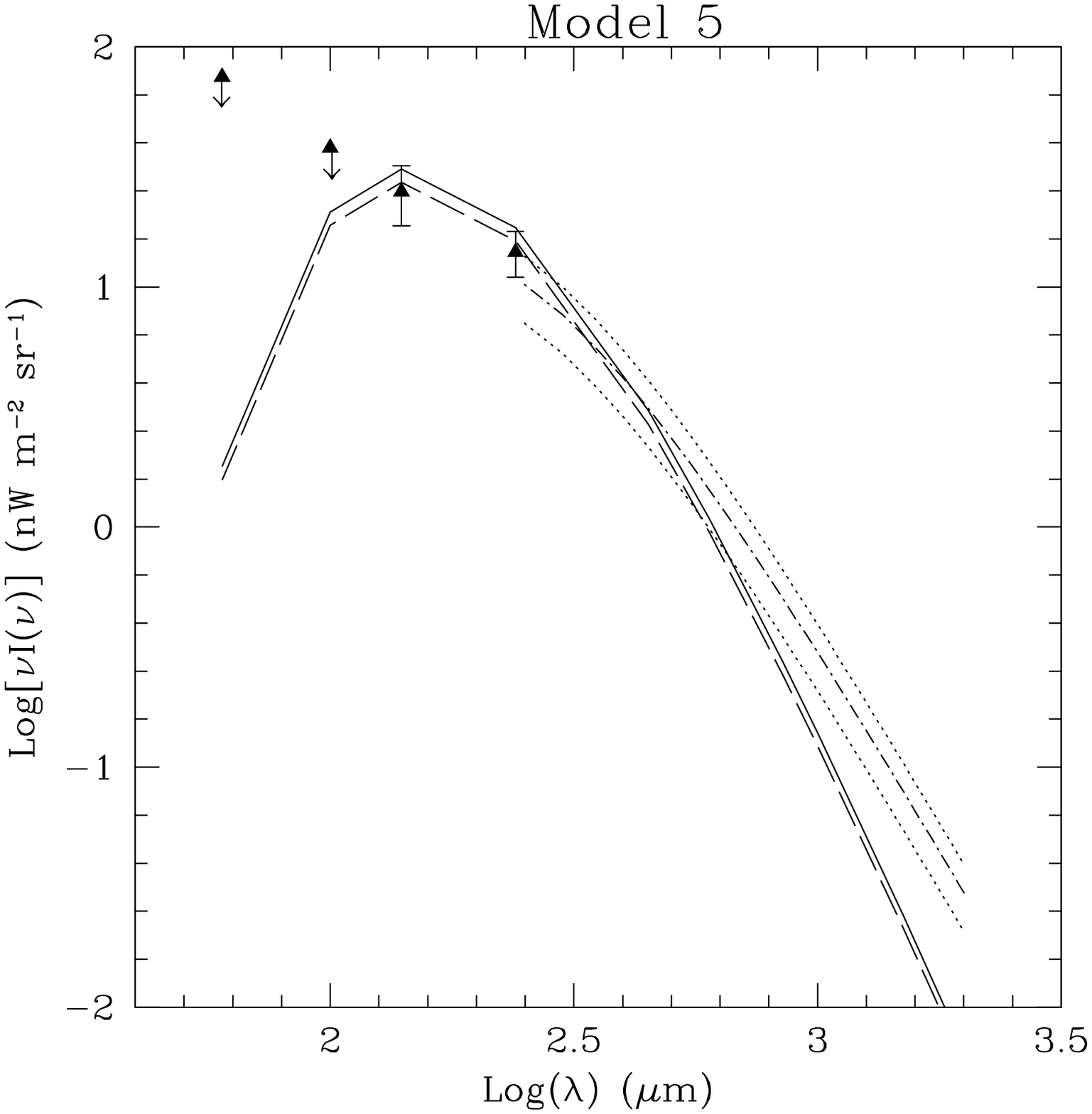}
\figcaption[ch_figure17.ps]{The FIR background predicted by the
SRF$_{int}$ of Model~5 is compared with the data from COBE DIRBE and
FIRAS. The symbols and lines are as in Figure~5.}
\end{figure}

\begin{figure}
\plotone{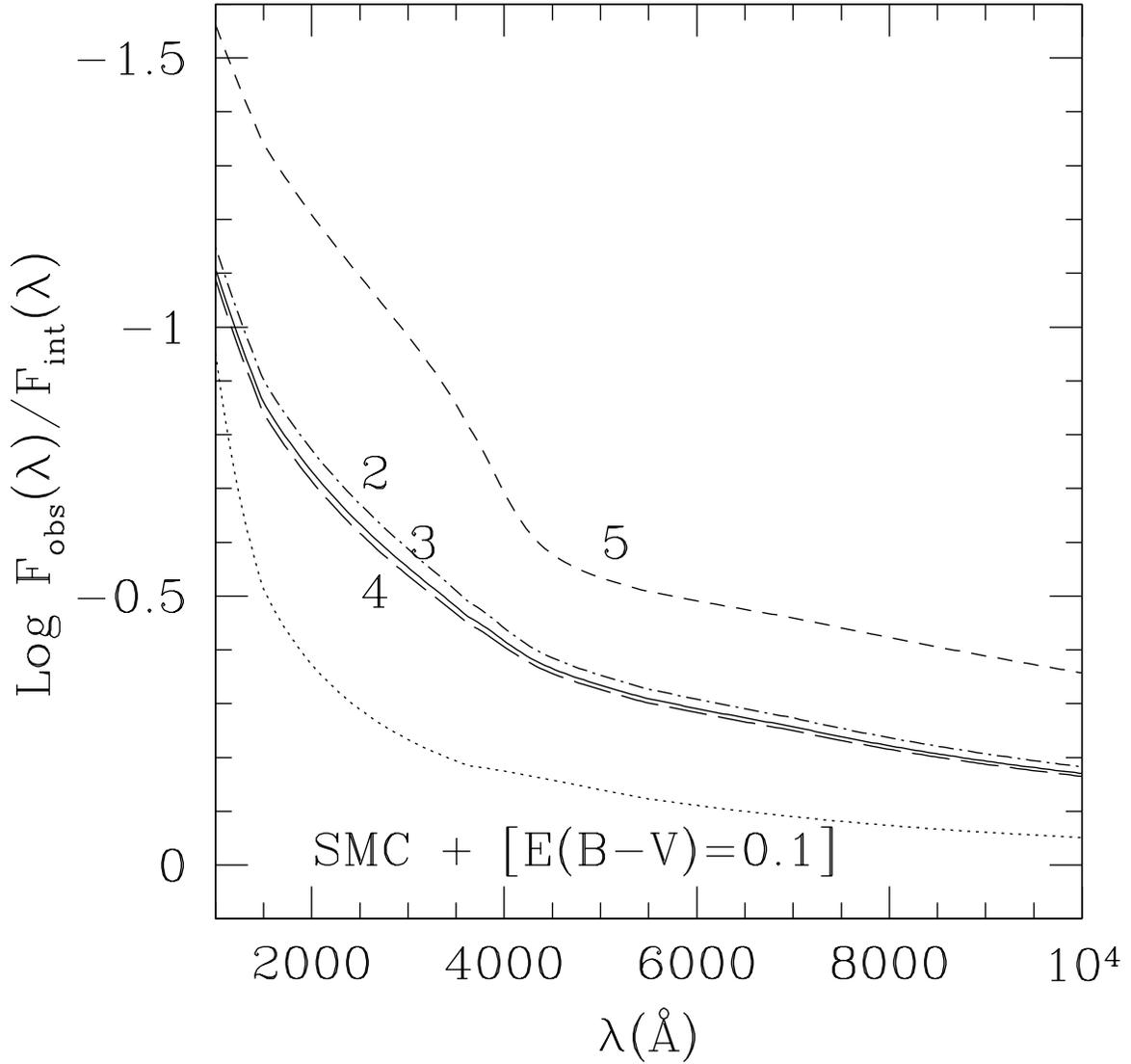}
\figcaption[ch_figure18.ps]{The relative and absolute attenuations
predicted by Models~2 to 5 in correspondence of their {\it peak dust
column densities} E(B$-$V) are plotted as a function of
wavelength. Symbols and line conventions are as in Figure~3. For
comparison, the attenuation produced by a foreground screen of dust
with E(B$-$V)=0.1 and an SMC curve is also shown. Models~2 to 4
produce relative attenuations which are comparable to the dust screen,
but absolute attenuations about 50\% higher. All Models are shown for the 
dust/star scaleheight ratio hh1.}
\end{figure}

\begin{figure}
\plotone{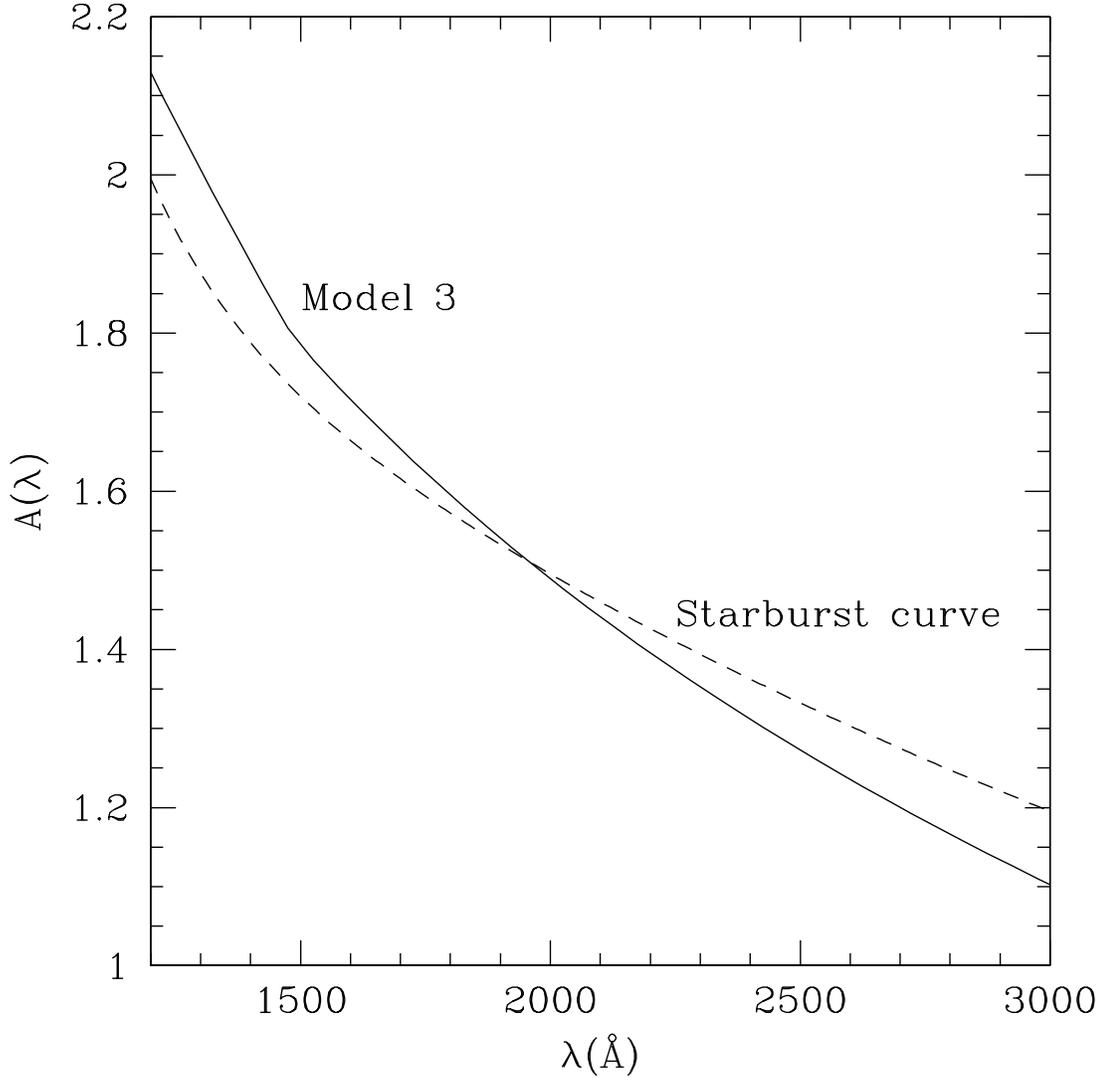}
\figcaption[ch_figure19.ps]{The attenuation A($\lambda$) of a galaxy at
z=3 is shown as a function of wavelength in the UV range for Model~3
(solid line) and the ratio h$_{dust}$/h$_{star}$ of hh1. For
comparison, the attenuation produced by the `starburst' reddening curve
(Calzetti et al. 1994, Calzetti 1997), rescaled to match the
attenuation from Model~3, is also shown (dashed line). The rescaling
factor is E(B$-$V)=0.154. The two attenuations are similar, with the
largest discrepancy being about 13\%~ in the range
1200--3000~\AA. Thus the starburst reddening curve can be used as an
estimator of galaxy opacity at high redshift, provided the appropriate
color excess E(B$-$V) is used.}
\end{figure}

\clearpage
\begin{figure}
\plotone{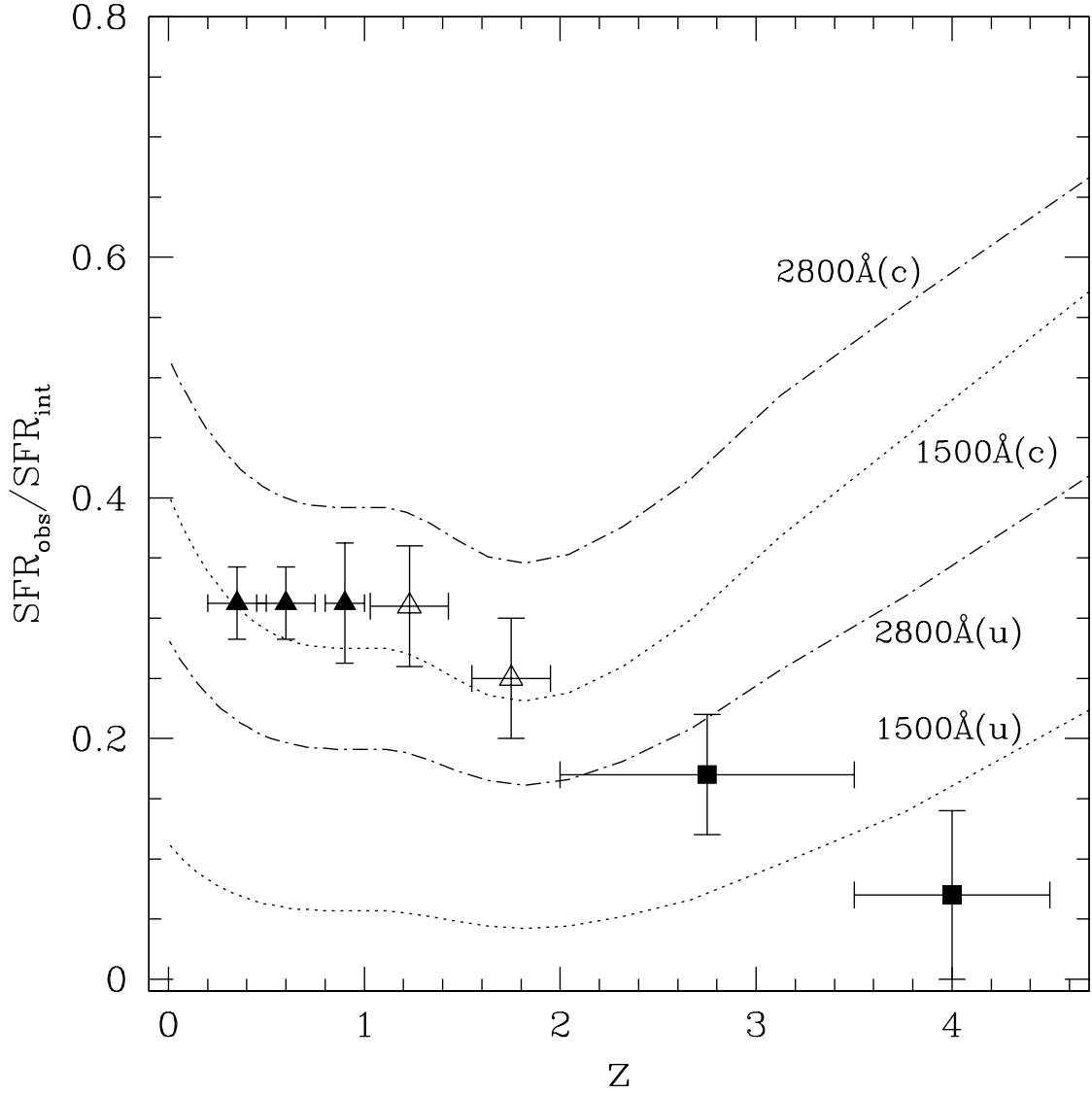}
\figcaption[ch_figure20.ps]{As Figure~10 (Model~3), for two different
geometries of the dust/star distribution: (u) the dust distribution is 
a uniform shell surrounding the stars; (c) the dust distribution
is a clumpy shell surrounding the stars. The emerging-to-total light
ratios are shown at both 2800~\AA~ (dash-dotted lines) and the
1500~\AA~ (dotted lines). The ratio h$_{dust}$/h$_{star}$ is given by
hh1.}
\end{figure}

\begin{figure}
\plotone{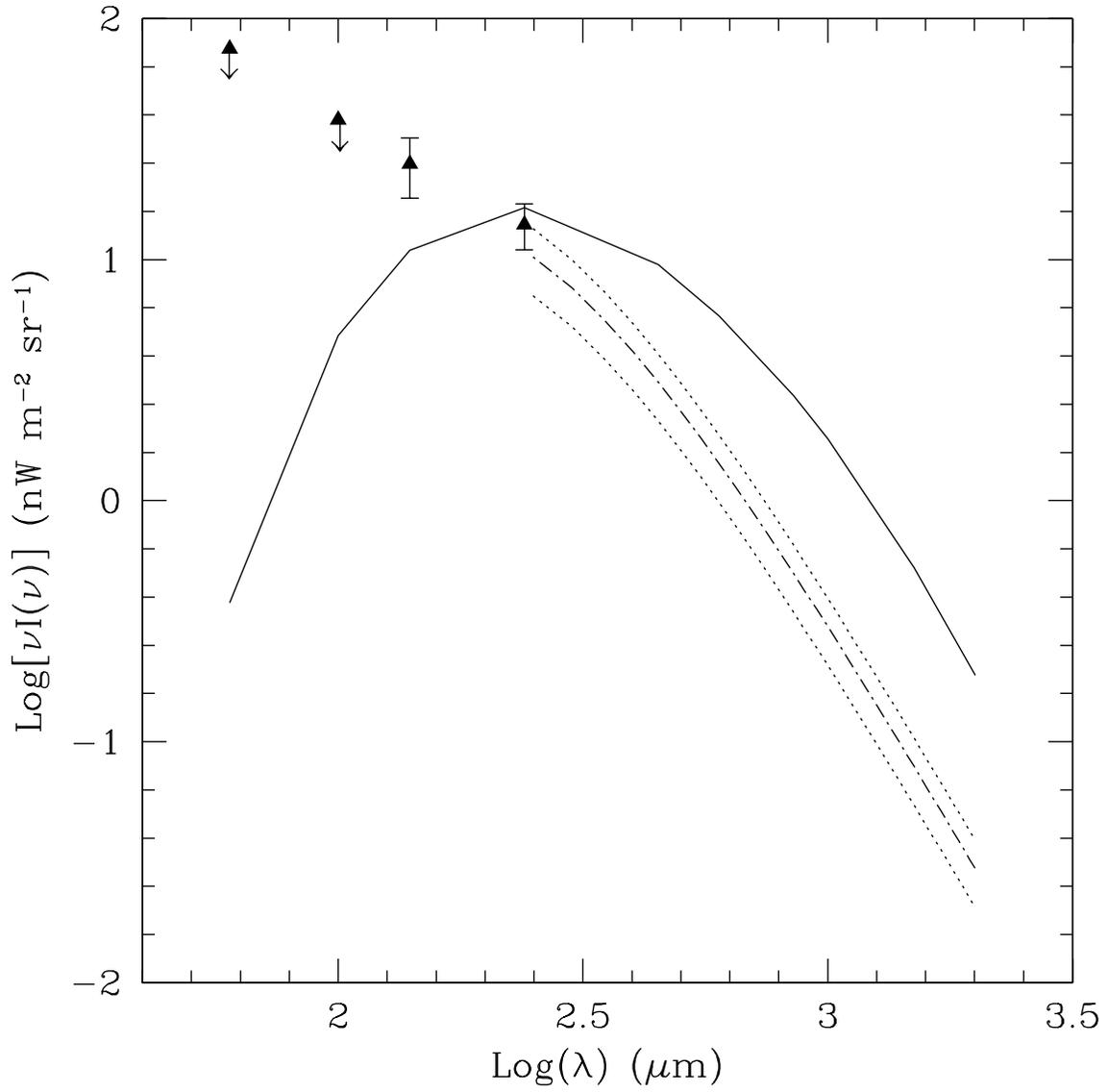}
\figcaption[ch_figure21.ps]{As Figure~11 (Model~3), for a constant temperature 
of the dust, T=21~K.}
\end{figure}

\begin{figure}
\plotone{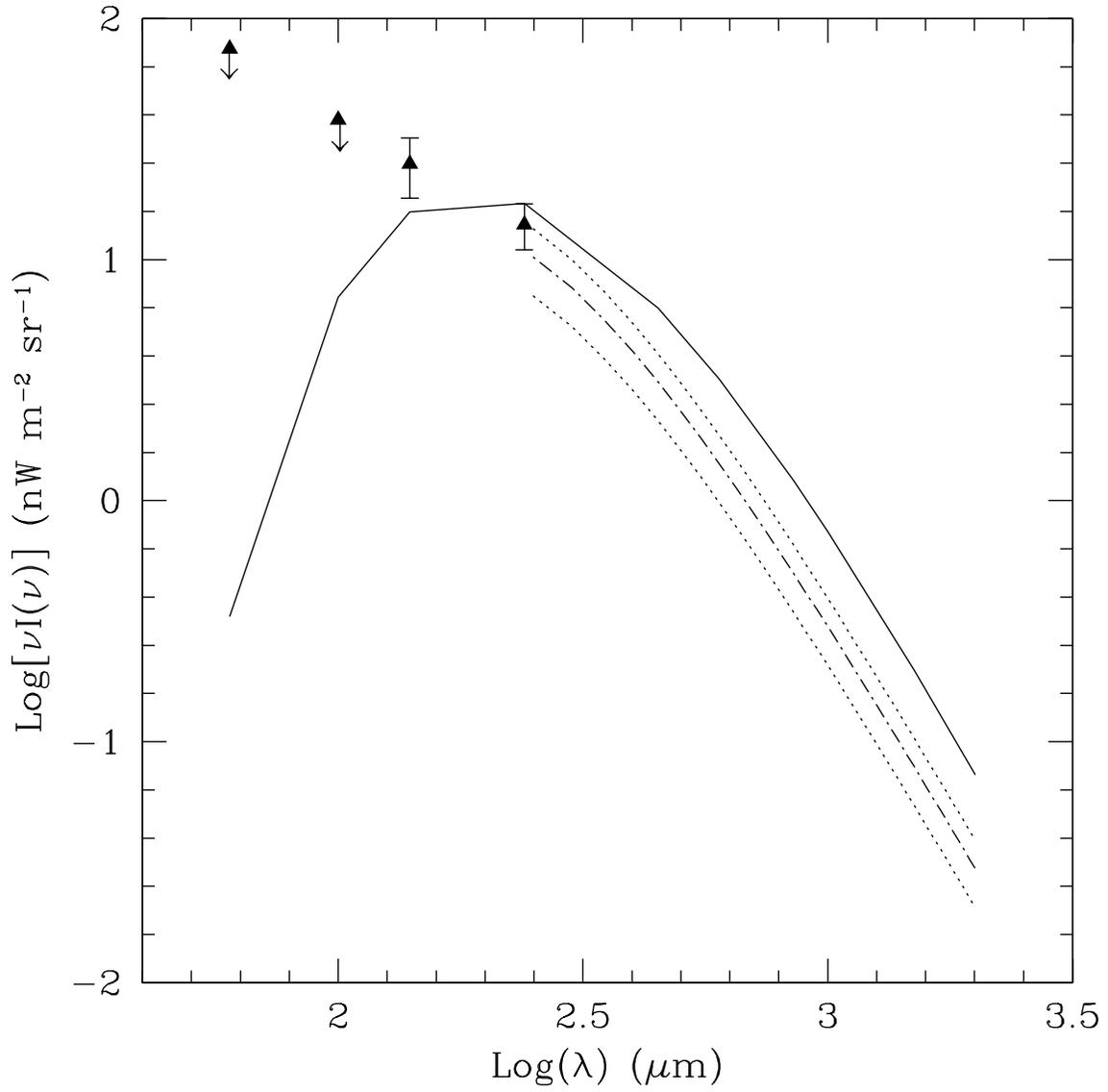}
\figcaption[ch_figure22.ps]{As Figure~11 (Model~3), for a dust emissivity 
index $\epsilon$=1.}
\end{figure}


\clearpage

\begin{thebibliography}{}
\bibitem[Armus et al. 1998a]{arm98a} Armus, L., Soifer, B.T., Murphy, T.W., 
Neugebauer, G., Evans, A.S., \& Matthews, K. 1998a, ApJ, 495, 276
\bibitem[Armus et al. 1998b]{arm98b} Armus, L., Matthews, K., 
Neugebauer, G., \& Soifer, B.T. 1998b, ApJ, in press (Letters)
\bibitem[Aragon-Salamanca et al. 1998]{ara98} Aragon-Salamanca, A., 
Baugh, C.M., \& Kauffmann, G. 1998, MNRAS, 297, 427
\bibitem[Barger et al. 1998]{bar98} Barger, A.J., Cowie, L.L., Sanders, 
D.B., Fulton, E., Taniguchi, Y., Sato, Y., Kawara, K. \& Okuda, H. 1998, 
Nature, embargoed (astroph/9806317)
\bibitem[Barlow \& Tytler 1998]{bar98} Barlow, T.A., \& Tytler, D. 1998, 
AJ, 115, 1725
\bibitem[Baugh, Cole \& Frenk 1996]{bau96} Baugh, C.M., Cole, S., \& Frenk, 
C.S. 1996, MNRAS, 283, 1361
\bibitem[Berlind et al. 1997]{ber97} Berlind, A.A., Quillen, A.C., 
Pogge, R.W., \& Sellgren, K. 1997, AJ, 114, 107
\bibitem[Bohlin, Savage \& Drake 1978]{boh78} Bohlin, R.C., Savage, B.D., 
\& Drake, J.F. 1978, ApJ, 224, 132
\bibitem[Boisse` et al. 1998]{boi98} Boiss\'e, P., Le Brun, V., 
Bergeron, J., \& Deharveng, J.-M. 1998, A\&A, 333, 841 
\bibitem[Bouchet et al. 1985]{bou85} Bouchet, P., Lequeux, J., Maurice, E., 
Prevot, L., \& Prevot-Burnichon, M.L. 1985, A\&A, 149, 330 
\bibitem[Bruzual \& Charlot 1996]{bru96}   Bruzual, G.~A., \& Charlot, 
S. 1996, private communication
\bibitem[Buat \& Burgarella 1998]{bua98} Buat, V., \& Burgarella, D. 1998, 
A\&A, in press
\bibitem[Calzetti 1997a]{cal97a}   Calzetti, D. 1997a, in The Ultraviolet 
Universe at Low and High Redshift: Probing the Progress of Galaxy Evolution, 
eds. W.H. Waller, M.N. Fanelli, J.E. Hollis \& A.C. Danks, AIP Conf. Proc. 
408 (Woodbury: AIP), 403
\bibitem[Calzetti 1997b]{cal97b}   Calzetti, D. 1997b, AJ, 113, 162
\bibitem[Calzetti et al. 1994]{cal94} Calzetti, D., Kinney, A.L., \& 
Storchi-Bergmann, T. 1994, ApJ, 429, 582
\bibitem[Calzetti et al. 1995]{cal95}   Calzetti, D., Bohlin, R.C., 
 Kinney, A.L., Storchi-Bergmann, T., \& Heckman, T.M. 1995, \apj, 443, 136
\bibitem[Cram 1998]{cra98} Cram, L.E. 1998, ApJ, in press (Letters) 
\bibitem[Connolly et al. 1997]{con97} Connolly, A.J., Szalay, A.S., 
Dickinson, M., Subbarao, M.U., \& Brunner, R.J. 1997, ApJ, 486, L11
\bibitem[Cox et al. 1986]{cox86} Cox, P., Kr\"ugel E., \& Mezger P.G. 1986, 
A\&A, 155, 380
\bibitem[Dehnen \& Binney 1998]{deh98} Dehnen, W., \& Binney, J. 1998, MNRAS, 
294, 429
\bibitem[D\'esert et al. 1990]{des90} D\'esert, F.-X., Boulanger, F., 
\& Puget, J.L. 1990, A\&A, 237, 215
\bibitem[Dey et al. 1995]{dey95} Dey, A., SPinrad, H., \& Dickinson, M. 1995, 
ApJ, 440, 515
\bibitem[Dickinson 1998]{dic98} Dickinson, M. 1998, in The Hubble 
Deep Field, STScI May Symposium, eds. M. Livio, S.M. Fall, \& P. Madau, 
(Cambridge: CUP), in press
\bibitem[Downes \& Solomon]{dow98} Downes, D., \& Solomon, P.M. 1998, ApJ, 
in press
\bibitem[Dwek 1998]{dwe98} Dwek, E. 1998, ApJ, 501, 643
\bibitem[Eales et al. 1998]{eal98} Eales, S., Lilly, S., Gear, W., Dunne, 
L., Bond, J.R., Hammer, F., Le F\`evre, O., \& Cramption, D. 1998, ApJ, 
submitted (Letters) 
\bibitem[Edmunds \& Phillips 1997]{edm97} Edmunds, M.G., \& Phillips, S. 1997, 
MNRAS, 733, 747
\bibitem[Fitzpatrick 1986]{fit86} Fitzpatrick, E.L. 1989, AJ, 92, 1068
\bibitem[Fixsen et al. 1998]{fix98} Fixsen, D.J., Dwek, E., Mather, J.C., 
Bennett, C.L., Shafer, R.A. 1998, ApJ, in press
\bibitem[Fukugita, Hogan \& Peebles 1998]{fuk98} Fukugita, M., Hogan, 
C.J., \& Peebles, P.J.E. 1998, ApJ, submitted
\bibitem[Gallego et al. 1995]{gal95} Gallego, J., Zamorano, J., 
Aragon-Salamanca, A., \& Rego, M. 1995, ApJ, 455, L1
\bibitem[Giavalisco et al. 1996]{gia96} Giavalisco, M., Steidel, C.C., 
\& Macchetto, F.M. 1996, ApJ, 470, 189
\bibitem[Giavalisco et al. 1998]{gia98} Giavalisco, M., et al. 1998, in 
prep.
\bibitem[Giovanelli et al. 1995]{gio95} Giovanelli, R., Haynes, M.P., Salzer, 
J.J., Wegner, G., Da Costa, L.N., \& Freudling, W. 1995, AJ 110, 1059
\bibitem[Gnedin \& Ostriker 1992]{gne92} Gnedin, N. Yu., \& Ostriker, J.P. 
1992, ApJ, 400, 1
\bibitem[Gonzalez et al. 1998]{gon98} Gonzalez, R.A., Allen, R.J., Dirsch, B., 
Ferguson, H.C., Calzetti, D., \& Panagia N. 1998, ApJ, in press
\bibitem[Gordon et al. 1997]{gor97} Gordon, K.D., Calzetti, D., \& Witt, 
A.N. 1997, ApJ, 487, 625
\bibitem[Gronwall 1998]{gro98} Gronwall, C., 
1998, in Dwarf Galaxies and Cosmology, the XXXIIIrd Recontres de
Moriond, eds. T.X. Thuan, C. Balkowski, V. Cayatte \& J. Tran Thanh
Van (Gif-sur-Yvette: Editions Fronti\'eres), in press
\bibitem[Hammer 1998]{ham98} Hammer, F. \& Flores, H. 1998,  in Dwarf 
Galaxies and Cosmology, the XXXIIIrd Recontres de Moriond, eds. T.X. Thuan, 
C. Balkowski, V. Cayatte \& J. Tran Thanh Van (Gif-sur-Yvette: Editions 
Fronti\'eres), in press (astroph/9806184)
\bibitem[Hauser et al. 1998]{hau98} Hauser, M.G., Arendt, R.G., Kelsall, T., 
Dwek, E., Odergard, N., Weiland, J.L., Freudenreich, H.T., Reach, W.T., 
Silverberg, R.F., Moseley, S.H., Pei, Y.C., Lubin, P., Mather, J.C., 
Shafer, R.A., Smoot, G.F., Weiss, R., Wilkison, D.T., \& Wright, E.L.  
1998, ApJ, accepted for publ.
\bibitem[Helou 1986]{hel86} Helou, G., 1986, ApJ, 311, L33
\bibitem[Hook et al. 1998]{hoo98} Hook, I.M., Shaver, P.A., \& McMahon, 
R.G. 1998, in The Young Universe: 
Galaxy Formation and Evolution at Intermediate and High Redshift, eds. 
S. D'Odorico, A. Fontana, \& E. Giallongo, ASP Conf. Series, in press
\bibitem[Hughes et al. 1998]{hug98} Hughes, D., Serjeant, S., Dunlop, J., 
Rowan-Robinson, M., Blain, A., Mann, R.G., Ivison, R., Peacock, J., 
Efstathiou, A., Gear, W., Oliver, S., Lawrence, A., Longair, M., 
Goldschmidt, P., \& Jenness, T. 1998, Nature, 394, 241
\bibitem[Hunter et al. 1996]{hun96} Hunter, D. A., O'Neil, Jr., E. J., 
Lynds, R., Shaya, E. J., et al. 1996, ApJ, 459, L27
\bibitem[Hunter et al. 1997]{hun97} Hunter, D. A.,  Light, R. M., 
Holtzman, J. A., Lynds, R., et al. 1997, ApJ, 478, 124
\bibitem[Ivison et al. 1998]{ivi98} Ivison, R.J., Smail, I., Le Borgne, J.-F., 
Blain, A.W., Kneib, J.-P., Bezecourt, J., Kerr, T.H., \& Davies, J.K. 1998, 
MNRAS, 298, 583
\bibitem[Jones et al. 1994]{jon94} Jones, A.P., Tielens, A.G.G.M., 
Hollenbach, D.J., \& McKee, C.F. 1994, ApJ, 433, 797
\bibitem[Kauffmann 1996]{kau96a} Kauffman, G. 1996, MNRAS, 281, 487
\bibitem[Kauffmann \& Charlot 1998]{kau98} Kauffmann, G., \& Charlot, S. 
1998, MNRAS, 294, 705
\bibitem[Kauffmann, Charlot \& White 1996]{kau96} Kauffmann, G., Charlot, S., 
\& White, S.D.M. 1996, MNRAS, 283, L117
\bibitem[Kennicutt 1998]{ken98} Kennicutt, R.C., 1998, ApJ, 498, 541
\bibitem[Kennicutt et al. 1994]{ken94} Kennicutt, R.C., Tamblyn, P., \& 
Congdon, C.E. 1994, ApJ, 435, 22
\bibitem[Kroupa 1995]{kro95} Kroupa, P. 1995, MNRAS 277, 1491
\bibitem[Krugel et al. 1998]{kru98} Krugel, E., Siebenmorgen, R., Zota, V., 
\& Chini, R. 1998, A\&A, 331, L9
\bibitem[Kunth et al. 1995]{kun95} Kunth, D., Matteucci, F., \& Marconi, 
G. 1995, A\&A, 297, 634
\bibitem[Kylafis \& Bahcall 1987]{kyl87} Kylafis, N.D., \& Bahcall. J.N. 
1987, ApJ, 317, 637
\bibitem[Larson 1998]{lar98} Larson, R.B. 1998, MNRAS, in press 
(astroph/9808145)
\bibitem[Lehnert \& Heckman 1996]{leh96} Lehnert, M.D., \& Heckman, T.M. 1996,
ApJ, 472, 546
\bibitem[Leitherer \& Heckman 1995]{lei95} Leitherer, C., \& Heckman, T.M. 
1995, ApJS, 96, 9
\bibitem[Lilly et al. 1998]{lil98} Lilly, S., Eales, S.A., Gear, W.K., 
Bond, J.R., Dunne, L., Hammer, F., Le F\`evre, O., \& Crampton, D. 1998, in 
NGST: Science and Technological Challenges, to be published by the European 
Space Agency
\bibitem[Lilly et al. 1996]{lil96} Lilly, S.J., Le F\'evre, O., Hammer, F. 
\& Crampton, D. 1996, ApJ, 460, L1
\bibitem[Lonsdale-Persson \& Helou 1987]{lon87} Lonsdale-Persson, C.J., 
\& Helou, G. 1987, ApJ, 314, 513
\bibitem[Lowenthal et al. 1997]{low97} Lowenthal, J., Koo, D., Guzman, R., 
Gallego, J., Phillips, A.C., Faber, S.M., Vogt, N.P., Illingworth, G.D., \& 
Gronwall, C. 1997, ApJ, 481, 673
\bibitem[Madau et al. 1996]{mad96} Madau, P., Ferguson, H.C., Dickinson, M.E., 
Giavalisco, M., Steidel, C.C., \& Fruchter, A. 1996, MNRAS, 283, 1388
\bibitem[Madau, Pozzetti, \& Dickinson 1998]{mad98} Madau, P., Pozzetti, 
L., \& Dickinson, M. 1998, ApJ, 498, 106 
\bibitem[Marzke et al. 1994]{mar94} Marzke, R.O., Huchra, J.P., \& Geller, 
M.J. 1994, ApJ, 428, 43
\bibitem[Massey et al. 1995]{mas95} Massey, P., Lang, C. C., DeGioia-Eastwood, 
K., \& Garmany, C. D. 1995, ApJ, 438, 188
\bibitem[Meurer et al. 1997]{meu97} Meurer, G. R., Heckman, T.M., 
Lehnert, M.D., Leitherer, C., \& Lowenthal, J. 1997, AJ, 114, 54
\bibitem[Meurer, Heckman, \& Calzetti]{meu98} Meurer, G. R., Heckman, T. M.,
\& Calzetti, D. 1998, in preparation
\bibitem[Moriondo et al. 1998]{mor98} Moriondo, G., Giovanelli, R., 
\& Haynes, M.P. 1998, A\&A, in press.
\bibitem[Mushotzky \& Loewenstein 1997]{mus97} Mushotzky, R.F., \& 
Loewenstein, M. 1997, ApJ, 481, L63
\bibitem[Omont et al. 1996]{omo96} Omont, A., McMahon, R.G., Cox, P., 
Kreysa, E., Bergeron, J., Pajot, F., \& Storrie-Lombardi, L.J. 1996, A\&A, 
315, 1
\bibitem[Ortolani et al. 1995]{ort95} Ortolani, S., Renzini, A., Gilmozzi, 
R., Marconi, G., Barbuy, B., Bica, E., \& Rich, M.R. 1995, Nature, 377, 701  
\bibitem[Pagel 1987]{pag87} Pagel, B.E.J. 1987, in The Galaxy, G. Gilmore 
\& B. Carswell eds. (Dordrecht: Reidel), 341
\bibitem[Pagel 1998]{pag98} Pagel, B.E.J. 1998, in Dwarf Galaxies and 
Cosmology, the XXXIIIrd Recontres de Moriond, eds. T.X. Thuan, C. Balkowski, 
V. Cayatte \& J. Tran Thanh Van (Gif-sur-Yvette: Editions Fronti\'eres), 
in press
\bibitem[Pei \& Fall 1995]{pei95} Pei, Y.C., \& Fall, S.M. 1995, ApJ, 454, 69
\bibitem[Pettini et al. 1998b]{pet98b} Pettini, M., Ellison, S.L., Steidel, 
C.C., \& Bowen, D.V. 1998b, ApJ, in press
\bibitem[Pettini et al. 1998a]{pet98a} Pettini, M., Kellogg, M., Steidel, 
C.C., Dickinson, M., Adelberger, K.L., \& Giavalisco, M. 1998a, ApJ, in press
\bibitem[Pettini et al. 1997a]{pet97a} Pettini, M., King, D.L., Smith, L.J., 
\& Hunstead, R.W. 1997a, ApJ, 478, 536
\bibitem[Pettini et al. 1997b]{pet97b} Pettini, M., Smith, L.J., King, D.L., 
\& Hunstead, R.W. 1997b, ApJ, 486, 665
\bibitem[Persic \& Salucci 1992]{per92} Persic, M., \& Salucci, P. 1992, 
MNRAS, 258, 14P
\bibitem[Phillips et al. 1995]{phi95} Phillips, A.C., Bershady, M.A., 
Forbes, D.A., Koo, D.C., Illingworth, G.D., Reitzel, D.B., Griffith, R.E., 
\& Windhorst, R.A. 1995, ApJ, 444, 21
\bibitem[Renzini 1997]{ren97} Renzini, A. 1997, ApJ, 488, 35 
\bibitem[Renzini 1998]{ren98} Renzini, A. 1998, in The Young Universe: 
Galaxy Formation and Evolution at Intermediate and High Redshift, eds. 
S. D'Odorico, A. Fontana, \& E. Giallongo, ASP Conf. Series, in press
\bibitem[Rowan-Robinson \& Crawford 1989]{row89} Rowan-Robinson, M., \& 
Crawford, J. 1989, MNRAS, 238, 523
\bibitem[Rowan-Robinson et al. 1997]{row97} Rowan-Robinson, M., et al. 1997, 
MNRAS, 289, 490
\bibitem[Sawicki \& Yee 1998]{saw98} Sawicki, M., \& Yee, H. K. C. 1998,
AJ, in press
\bibitem[Scalo 1998]{sca98} Scalo, J. M. 1998, in Proc. 38th Herstmonceaux 
Conf. on the Stellar IMF, ed. Gilmore, Parry, \& Ryan, in press.
\bibitem[Schramm \& Turner 1998]{sch98} Schramm, D.N., \& Turner, M.S. 
1998, Rev. Mod. Phys., 70, 303
\bibitem[Schechter \& Dressler 1987]{sch87} Schechter, P.L., \& Dressler, A. 
1987, AJ, 94, 563
\bibitem[Scoville et al. 1998]{sco98} Scoville, N.Z., Evans, A.S., 
Dinshaw, N., Thompspn, R., Rieke, M., Schneider, G., Low, F.J., Hines, D., 
Stobie, B., Becklin, E., \& Hepps, H. 1998, ApJ (Letters), in press
\bibitem[Shaver et al. 1998]{sha98} Shaver, P.A., Hook, I.M., Jackson, C.A., 
Wall, J.V., Kellermann, K.I. 1998, in Highly Redshifted Radio Lines, eds. 
C. Carilli, S. Radford, K. Menten, \& G. Langston (San Francisco: PASP), 
in press
\bibitem[Shaver et al. 1996]{sha96} Shaver, P.A., Wall, J.V., Kellermann, 
K.I., Jackson, C.A., \& Hawkins, M.R.S. 1996, Nature, 384, 439
\bibitem[Sirianni et al. 1998]{sir98} Sirianni, M., Nota, A., Leitherer, C., 
De Marchi, G., \& Clampin, M. 1998, in prep.
\bibitem[Smail et al. 1997]{sma97} Smail, I., Ivison, R.J., \& Blain, A.W. 
1997, ApJ, 490, L5
\bibitem[Soifer \& Neugebauer 1991]{soi91} Soifer, B.T., \& Neugebauer, G. 
1991, AJ, 101, 354
\bibitem[Steidel et al. 1998]{ste98} Steidel, C.C., Adelberger, K.L.,
Giavalisco, M., Dickinson, M., Pettini, M., \& Kellogg, M. 1998,
proceedings of the Royal Society Discussion Meeting, Phil. Trans. R. Soc. 
Lond., in preparation
\bibitem[Steidel et al. 1996]{ste96} Steidel, C.C., Giavalisco, M.,
 Pettini, M., Dickinson, M., \& Adelberger, K.L. 1996, ApJ, 462, L17
\bibitem[Storrie-Lombardi et al. 1996]{sto96} Storrie-Lombardi, L.J., 
McMahon, R.G., \& Irwin, M.J. 1996, MNRAS, 283, L79
\bibitem[Tielens 1990]{tie90} Tielens, A.G.G.M. 1990, in Carbon in the 
Galaxy, ed. J.C. Tarter, S. Chang, \& D. De Frees (NASA CP 3061), 59
\bibitem[Tosi 1998]{tos98} Tosi, M. 1998, in Dwarf Galaxies and 
Cosmology, the XXXIIIrd Recontres de Moriond, eds. T.X. Thuan, C. Balkowski, 
V. Cayatte \& J. Tran Thanh Van (Gif-sur-Yvette: Editions Fronti\'eres), 
in press
\bibitem[Treyer et al. 1998]{tre98} Treyer, M.A., Ellis, R.S., Milliard, B., 
Donas, J., \& Bridges, T.J. 1998, MNRAS, in press 
\bibitem[Trewhella 1997]{tre97} Trewhella, M.  1997, in The Ultraviolet 
Universe at Low and High Redshift: Probing the Progress of Galaxy Evolution, 
eds. W.H. Waller, M.N. Fanelli, J.E. Hollis \& A.C. Danks, AIP Conf. Proc. 
408 (Woodbury: AIP), 349
\bibitem[Villumsem \& Strauss 1987]{vil87} Villumsen, J.V., \& Strauss, 
M.A. 1987, ApJ, 322, 37
\bibitem[Vladilo 1998]{vla98} Vladilo, G. 1998, ApJ, 493, 583
\bibitem[Wang \& Heckman 1996]{wan96} Wang, B., \& Heckman, T.M. 1996, ApJ, 
457, 645
\bibitem[Webster et al. 1995]{web95} Webster, R.L., Francis, P.J., 
Peterson, B.A., Drinkwater, M.J., \& Masci, F.J. 1995, Nature, 375, 469
\bibitem[White \& Frenk 1991]{whi91} White, S.D.M., \& Frenk, C.S. 1991, 
ApJ, 379, 25 
\bibitem[White, Keel \& Conselice 1996]{whi96} White, R.E., Keel, W.C., 
\& Conselice, C.J. 1996, pre-print (astroph/9604029)
\bibitem[Williams et al. 1996]{wil96} Williams, R.E., et al. 1996, AJ, 112, 
1335
\bibitem[Witt \& Gordon 1996]{wit96} Witt, A.N., \& Gordon, K.D. 1996, ApJ, 
463, 681
\bibitem[Witt et al. 1992]{wit92} Witt, A.N., Thronson, H.A., \& Capuano, 
J.M. 1992, ApJ, 393, 611
\bibitem[Worthey 1994]{wor94} Worthey, G. 1994, ApJS, 95, 107
\bibitem[Xu \& Buat 1995]{xu95} Xu, C., \& Buat, V. 1995, A\&A, 293, L65
\bibitem[Yamashita 1992]{yam92} Yamashita, K. 1992, in Frontiers of X-ray 
Astronomy, eds. Y. Tanaka \& K. Koyama (Tokyo: Universal Academy Press), 475
\bibitem[Young et al. 1989]{you89} Young, J.S., Xie, S., Kenney, J.D.P., \& 
Rice, W.L. 1989, ApJS, 70, 699. 
\end{thebibliography}
\end{document}